\documentclass[iop,apjl,numberedappendix]{emulateapj}
\usepackage{amssymb}
\usepackage{amsmath}
\usepackage{graphicx}
\usepackage{natbib}
\usepackage{url}
\usepackage{paralist}
\usepackage{epsfig}

\def\h18{\hbox{H1821$+$643\,}}

\slugcomment{ }

\shorttitle{Suppression of AGN Turbulence by Magnetic Fields}
\shortauthors{C.~J.~Bambic et al.}

\begin{document}

\title{Suppression of AGN-Driven Turbulence by Magnetic Fields in a Magnetohydrodynamic Model of the Intracluster Medium}

\author{Christopher~J.~Bambic\altaffilmark{1}, Brian~J.~Morsony\altaffilmark{1}, and Christopher~S.~Reynolds\altaffilmark{1,2,3} }

\altaffiltext{1}{Department of Astronomy, University of Maryland, College Park, MD 20742-2421, USA; cbambic@umd.edu}
\altaffiltext{2}{Joint Space-Science Institute (JSI), College Park, MD 20742-2421, USA}
\altaffiltext{3}{Institute of Astronomy, Madingley Road, CB3 0HA Cambridge, United Kingdom}

\begin{abstract}
\noindent We investigate the role of AGN feedback in turbulent heating of galaxy clusters. Specifically, we analyze the production of turbulence by g-modes generated by the supersonic expansion and buoyant rise of AGN-driven bubbles. Previous work which neglects magnetic fields has shown that this process is inefficient, with less than 1\% of the injected energy ending up in turbulence. This inefficiency is primarily due to the fact that the bubbles are shredded apart by hydrodynamic instabilities before they can excite sufficiently strong g-modes. Using a plane-parallel model of the ICM and 3D ideal MHD simulations, we examine the role of a large-scale magnetic field which is able to drape around these rising bubbles, preserving them from hydrodynamic instabilities. We find that, while magnetic draping appears better able to preserve AGN-driven bubbles, the driving of g-modes and the resulting production of turbulence is still inefficient. The magnetic tension force prevents g-modes from transitioning into the nonlinear regime, suppressing turbulence in our model ICM. Our work highlights the ways in which ideal MHD is an insufficient description for the cluster feedback process, and we discuss future work such as the inclusion of anisotropic viscosity as a means of simulating high $\beta$ plasma kinetic effects. These results suggest the hypothesis that other mechanisms of heating the ICM plasma such as sound waves or cosmic rays may be responsible for observed feedback in galaxy clusters.
 
\end{abstract}



\section{Introduction}\label{intro}

Atmospheric gas dynamics is a ubiquitous problem in astrophysics, determining the evolution of systems from exoplanet atmospheres to the intracluster medium (ICM) of galaxy clusters. Early works by \cite{Schwarzschild} and \cite{Jeans} studied the stability of atmospheres using linear fluid theory, yet nonlinear physics remained inaccessible. With developments in computational power, we are now able to probe the nonlinear physics relevant to magnetized, turbulent atmospheres. While this physics is relevant to a number of astrophysical systems, we will focus primarily on the hot ($T$ $\sim$ 10$^7$ - 10$^8$ K), diffuse ($n$ $\sim$ 10$^{-2}$ - 10$^{-3}$ cm$^{-3}$) ICM present within clusters of galaxies.

Galaxy clusters are the most massive baryonic structures in the universe. While most of their mass ($ \sim$ 84\%) lies in dark matter, $\sim$ 12\% of the mass is present within a hot atmosphere of plasma held in hydrostatic equilibrium by the dark matter potential of the cluster \citep{Felten1966, Parrish2012}. Without any energy input, this plasma would radiatively cool in $\sim$ 1 Gyr, leading to a so-called ``cooling catastrophe'' with high star formation rates and massive central cluster galaxies unobserved in nature \citep{Fabian1994}. Instead of this catastrophe, the ICM atmosphere appears to be in an approximate thermal equilibrium \citep{Peterson2006}. Given that models based on conduction have failed to account for this observed equilibrium \citep{Binney1981, Narayan2001, Ruszkowski2002, Kim2003, Zakamska2003, Voigt2004, Balbus2008, Bogdanovic2009, Yang2016a}, it is now widely accepted that feedback from a central active galactic nucleus (AGN) located in the brightest cluster galaxy (BCG) is responsible for the energy input offsetting radiative cooling \citep{Reynolds2002, Churazov2000, Churazov2002}.

AGN feedback is well-established observationally (see \cite{Fabian2012} and references therein for a more complete discussion). Within galaxy clusters, feedback occurs in the kinetic or radio-mode where a jet from the central AGN provides the mechanical energy necessary to offset cooling. Approximately 70\% of cluster BCGs are radio-loud \citep{Burns1990}, indicating the presence of these jets. In addition, X-ray observations have revealed clear cavities and bubbles blown out by the AGN \citep{Fabian2000} with cavity energies that correlate well with the cooling/radiative losses, indicating that central jetted AGN are in fact responsible for the observed thermal equilibrium.

Observations have revealed that there is enough energy in the jet to offset radiative cooling in the cluster \textit{if} the jet can heat the ICM efficiently. One might expect that energy is transferred primarily through shock heating driven by the supersonic expansion of hot plasma bubbles blown by the jets, yet these strong shocks are not observed in X-ray images of nearby clusters. The absence of these shocks indicates that the strong shock phase is short-lived and therefore does not provide the gradual, uniform heating necessary to account for the temperature profiles measured in galaxy clusters. Measurements of the Perseus Cluster with the Hitomi Soft X-ray Spectrometer (SXS) find a velocity dispersion measure of $\sigma \sim 164$ km/s \citep{Hitomi2016}. If this velocity dispersion is interpreted as turbulent motion, this measurement indicates a large-scale turbulent energy of approximately 4\% of the thermal energy. If this energy is transferred to small scales via a turbulent cascade and entirely dissipated as heat, radiative cooling can be offset and the cluster can remain in the observed thermal equilibrium provided the turbulent energy can be replenished sufficiently rapidly.

\cite{RBS15} (hereafter RBS) investigated the production of turbulence by g-modes (buoyancy modes) generated by the supersonic expansion and buoyant rise of AGN-driven bubbles in a plane-parallel model of the ICM using 3D hydrodynamics simulations. They found that the turbulent production was inefficient, with less than 1\% of the injected energy ending up in turbulent motions. The rising bubbles were unstable to the Kelvin-Helmholtz instability at their contact interface, and because the bubble material is less dense than the surrounding atmosphere, the bubbles were unstable to both the Rayleigh-Taylor and Richtmyer-Meshkov instabilities \citep{Richtmyer1960, Meshkov1969, Brouillette2002, Friedman2012}. These instabilities acted to shred apart the bubbles before they could rise high enough to excite g-modes sufficiently strong enough to induce the turbulence required to offset radiative cooling. Rather than going into incompressible g-modes, most of the energy was carried off in compressible sound waves or p-modes. RBS analyzed these processes using an idealized atmosphere (i.e. a plane-parallel background with periodic boundary conditions in the horizontal dimensions), but recent work employing more realistic cluster geometries has reached similar conclusions regarding the inefficiency with which AGN jets drive turbulence \citep{Yang2016, Weinberger2017, Hillel2016, Hillel2017, Bourne2017}.

In this work, we extend the work of RBS to include the effects of magnetic fields. The ICM is a hot, ionized plasma. As such, magnetic fields generated either through some seed or dynamo mechanism are likely to be present within the plasma. Indeed, Faraday rotation measurements have revealed the presence of large-scale magnetic fields in clusters \citep{Bonafede2010}; however, these fields are weak, with a typical ratio between the thermal and magnetic pressures (the plasma $\beta$ parameter) around 100. Even though these fields are weak, they can ``drape'' around rising bubbles, acting to suppress the hydrodynamic instabilities which shredded the bubbles in the RBS simulations \citep{Dursi2008}.

The basic physics of magnetic draping is well-established. In a perfectly conducting plasma, Alfv\`en's theorem states that the magnetic field lines are ``frozen'' into the plasma, meaning that the motion of the plasma induces motion of the magnetic field and vice-versa. The bubble draws up magnetic field at the front interface as it rises, forming a layer of increased magnetic field which is ``draping'' around the bubble. This draping layer can act to suppress the growth of fluid instabilities at small scales, preserving the bubble from being shredded apart and allowing the bubble to rise higher into the atmosphere.

The bubble will rise in the atmosphere until the bubble entropy equals the entropy of the surrounding medium. This process is largely mediated by the mixing of bubble plasma with the ICM and is sensitive to shredding of the bubble interface through various instabilities. By including an effective surface tension to the bubble through magnetic draping, the entropy of the bubble is kept nominal, allowing a higher rise in the atmosphere and suppressing mixing between the bubble and ICM plasmas.

The upward rise of the bubble displaces ICM plasma on large scales. Gravity acts to restore the displaced plasma, inducing a buoyancy oscillation or ``g-mode'' which acts as a sloshing of cluster plasma. The g-mode oscillates at a frequency defined by \citep{Balbus2010},
\begin{equation}
	\Omega_g^2 = (v_A k_{\parallel})^2 + \frac{k_h^2}{k^2} N^2,
\end{equation}
where $k_h$ is the wavenumber $\textbf{k}$ projected on the horizontal plane, $k_{\parallel}$ the wavenumber projected along the magnetic field direction, $v_A$ the magnitude of the Alfv\`en velocity given by,
\begin{equation}
	\textbf{v}_A = \frac{\textbf{B}}{\sqrt{4 \pi \rho}},
\end{equation}
 and $N$ is the Brunt-V{\"a}is{\"a}l{\"a} frequency,
\begin{equation}
	N^2 = \frac{g(z)}{\gamma} \frac{\partial}{\partial z} \ln(P \rho^{-\gamma}).
\end{equation}
In Equations 2 and 3, $\textbf{B}$ is the magnetic field, $\rho$ is the gas density, $P$ is the gas pressure, $g(z)$ the gravitational acceleration, and $\gamma$ the adiabatic index of the gas. If a magnetic field can drape around the rising bubbles, preserving them from hydrodynamic instabilities and allowing the bubbles to rise higher, g-modes may be able to be produced which are strong enough to account for the turbulent energy in galaxy clusters.

In this paper, we investigate the role of magnetic fields on the production and development of AGN-driven turbulence. We set up a toy model of the ICM atmosphere, driving the atmosphere with thermal ``bombs'' or over-pressurizations which blow bubbles in the atmosphere. We then study the rising of the bubbles, their interaction with the magnetic fields through draping, the launching of g-modes, and the subsequent nonlinear decay of g-modes into turbulence. Finally, we measure the resulting efficiency of turbulent energy production and the effect of the magnetic fields on the suppression of this turbulence. 

In general, we find that the presence of magnetic fields does not allow driving of noticeably stronger g-modes. In addition, the magnetic tension force acts to halt the transition of the g-modes to turbulence, suppressing turbulence at small scales. \cite{Fabian2017} argued that g-modes could not account for the heating in the Perseus Cluster because the internal waves could not propagate radially to the cooling radius of the cluster within the cooling time. Our work indicates that even if g-modes could fill the volume of the cluster within the cooling time, large-scale magnetic fields in the cluster could act to suppress their nonlinear evolution to turbulence. Note, these results are unique to ideal magnetohydrodynamics (MHD) and may be subject to our launching mechanism for g-modes. Physics beyond ideal MHD and further study of the nonlinear interactions of buoyancy waves in a magnetized plasma are thus necessary to properly treat the problem of ICM plasma turbulence. In addition, other mechanisms for heating the ICM plasma such as sound waves (magnetosonic waves in the case of the ICM plasma) or cosmic rays may be responsible for the observed feedback in galaxy clusters.

This paper is organized as follows. In Section 2 we will discuss the necessary theoretical concepts including the effect of magnetic fields on suppression of instabilities, the transition of g-modes to turbulence, and the evolution of the turbulent cascade in ideal MHD. In Section 3 we will discuss our simulation setup for 2 separate simulations. In one simulation, we test a unidirectional magnetic field with a strength given by $\beta$ of 100. In the other, we explore the role of a helical field in preserving the bubble. In Section 4, we will present our results regarding magnetic draping and the role of field geometry on this process. In Section 5, we discuss the role of magnetic tension in suppressing turbulence and how we determined AGN-driven g-mode turbulence to be an inefficient mechanism for heating the ICM within the ideal MHD framework. In Section 6, we discuss the implications of our results and how they relate to alternative theories of heating the ICM. Finally, we conclude in Section 7.

\section{Theoretical Preliminaries}\label{theory}

The ICM is a hot, ionized, weakly-collisional, magnetized plasma. Therefore, our problem is inherently a plasma physics problem. In addition, the problem is highly anisotropic, with gravitational and magnetic fields providing preferential directions for the development of instabilities and turbulent eddies. Because of the large ion mean free path and the resulting weakly collisional nature of the plasma, a fully rigorous treatment of the problem requires kinetic plasma theory; however, because clusters are $\sim$ Mpc in diameter, our simulations are unable to resolve the scales where kinetic theory becomes relevant. Clusters are much larger than the ion gyroscale and AGN feedback occurs throughout cosmic time. As such, we undertake our study of the ICM using ideal MHD, although we will discuss some issues arising from weak collisionality in Section 6. 

A consequence of including magnetic fields in our simulation is the formation of a draping layer which can act to suppress the growth of hydrodynamic instabilities. The major instabilities we are interested in are the Kelvin-Helmholtz instability (KHI) which acts at the bubble-ICM interface and is driven by the rising of the bubble, the Rayleigh-Taylor instability (RTI) which results from the density gradient at the interface of the bubble and ICM, and the Richtmyer-Meshkov instability (RMI) which results from the upward rise of the bubble and the resulting sound wave reflection at the bubble-ICM interface. The KHI is stabilized by a uniform magnetic field acting parallel to the bubble-atmosphere interface if the speed of the bubble relative to the atmosphere does not exceed the root mean square Alfv\`en speed between the two media \citep{Chandrasekhar1961}. For parameters relevant to our problem, the bubble terminal velocity never exceeds the Alfv\`en speed. Thus, within our model system, the Kelvin-Helmholtz instability is suppressed by the magnetic draping layer.

The inclusion of a magnetic field at the bubble interface acts as an effective surface tension at the interface, causing a decay rather than a growth in a perturbation's amplitude for arbitrary $k$. We can describe the the RTI in the linear regime by the dispersion relation \citep{Chandrasekhar1961},
\begin{equation}
	\Gamma^2_{mRTI} = gk\Bigg( \frac{\rho_{ICM}-\rho_{bub}}{\rho_{ICM}+\rho_{bub}} - \frac{B^2 k_{\parallel}^2}{2 \pi (\rho_{ICM} + \rho_{bub}) gk} \Bigg),
\end{equation}
where $g$ is the gravitational acceleration, $B$ is the magnitude of the magnetic field, $\rho_{bub}$ is the bubble density, $\rho_{ICM}$ is the atmospheric density, $k$ is the magnitude of the wavenumber vector $\textbf{k}$, $\Gamma_{mRTI}$ is the growth rate, and $k_{\parallel}$ is the wavenumber along the initial magnetic field direction. Note that the magnetic field direction will change as the field is draped about the bubble.

The bubble interface is unstable to small perturbations when $\rho_{ICM} > \rho_{bub}$ in the range of $0 < k_{\parallel} < k_c$; however, the interface is stabilized for all $k_{\parallel} > k_c$, where $k_c$ is given by
\begin{equation}
	k_c^2 = \frac{2 \pi (\rho_{ICM} - \rho_{bub}) gk}{B^2}.
\end{equation}
The above equation indicates that the range of unstable wavenumbers decreases with increasing magnetic field strength. Note, however, that when $k_{\parallel} = 0$, i.e. the perturbation is perpendicular to the initial magnetic field, the interface is unstable to perturbations for all $k$. Therefore, the magnetic field acts only to stabilize the bubble in directions along the field lines, and any perturbations across them are unaffected by the field. This feature will result in an anisotropic development of the RTI referred to in this paper as the magnetic RTI which will be relevant in the results for Simulation 1 in Section 4. Because the RMI is an accelerated form of the RTI, we expect the same effect would be relevant for the RMI's evolution. The activation of these instabilities will be discussed further in our results for Simulation 1 and will serve as the primary motivation for the inclusion of a helical magnetic field (Simulation 2).

Next, we address the production of turbulence from these rising bubbles. In hydrodynamics, turbulence is excited by the ``stirring'' of a fluid, producing large eddies which in turn induce smaller and smaller eddies, eventually transferring energy from large to small scales in a turbulent cascade. This stirring which occurs on some large driving scale $L$ produces a rotation in the fluid velocity field, correlating to the production of a vorticity $\omega$ given by, 
\[
	\omega = \nabla \times \textbf{v}.
\]
In ideal MHD, this picture is modified by the presence of a magnetic field. In this case, colliding Alfv\`en wave packets transfer energy from large to small scales \citep{Howes2013}. Fundamentally, the dynamics resulting from these collisions is characterized by the vorticity of the magnetized fluid.

The vorticity evolution equation in ideal MHD for an incompressible fluid ($\nabla \cdot \textbf{v} = 0$) is given by,
\begin{equation}
\begin{split}
	\frac{D \omega}{Dt} = (\omega \cdot \nabla) \textbf{v} - (\nabla \cdot \textbf{v}) \omega + \frac{1}{\rho^2} \nabla \rho \times \nabla \Big(P + \frac{|\textbf{B}|^2}{2} \Big) \\
	 - \frac{1}{\rho^2} \nabla \rho \times \frac{(\textbf{B} \cdot \nabla) \textbf{B}}{4 \pi},
\end{split}
\end{equation}
where $D/Dt = \partial/\partial t + (\textbf{v}\cdot\nabla)$ is the convective derivative. 

G-modes are large-scale waves in a fluid where gravity acts as the restoring force. These waves result both in a change between iso-surfaces of density and pressure and in the motion of the magnetic field through flux freezing. A linear treatment shows that g-modes in a magnetized fluid are associated with vorticity in the horizontal plane. In our model, we choose a Cartesian geometry with gravity in the $-z$-direction. Therefore, linear g-modes are related to $\omega_x$ and $\omega_y$ components of vorticity.

When the nonlinear terms of the MHD equations become relevant, a coupling between wavenumbers allows for the decay of g-modes to turbulence. This ``breaking'' of the g-mode is likely due to a combination of density inversion within the wave, inducing a form of the RTI, as well as a parasitic KHI which develops from the shearing of the wave with the surrounding medium. The interaction of g-modes will result in a turbulent decay as long as the wave amplitude is sufficient that the nonlinear interaction timescale is much shorter than the buoyancy time. In other words, g-modes break when $(v_h k_h)^{-1}$ $\ll$ $N^{-1}$, where $v_h$ and $k_h$ are characteristic velocities and wavenumbers in the horizontal ($x$-$y$) plane and $N$ is the Brunt-V{\"a}is{\"a}l{\"a} frequency. We will use power spectra to probe this nonlinear interaction in Section 5, demonstrating that the field geometry plays a fundamental role in suppressing the nonlinear interaction which breaks the g-modes and produces turbulence.

In isotropic, homogeneous, hydrodynamic turbulence, the inertial range of the turbulent cascade is well-described by the Kolmogorov (1941) relation,
\[
	E(k) \propto \varepsilon^{2/3} k^{-5/3},
\]
where $\varepsilon$ is the energy dissipation rate per unit mass. This relation is arrived at through the assumption of locality of interactions in $k$-space. In MHD, this assumption is invalid since scales are coupled through the presence of magnetic fields and locality is violated. In addition, as was discussed before, magnetic fields and the gravitational field induce an anisotropy in the development of turbulence, violating the isotropy assumption. We will address these issues through the use of anisotropic power spectra, i.e. power spectra taken along and across the magnetic field separately.

In ideal MHD in the presence of a strong guide field, inertial range dynamics appear well-described by the critical balance conjecture, a statement of a balance between the Alfv\`en timescale and nonlinear interaction timescale. Stated mathematically,
\begin{equation}
	\omega_{MHD} \sim v_A k_{\parallel} \sim v_{\bot} k_{\bot},
\end{equation}
where $\omega_{MHD}$ is a typical fluctuation frequency, $v_{\bot}$ is a fluctuation perpendicular to the magnetic field, and $k_{\bot}$ is the perpendicular wavenumber \citep{Goldreich1995}. The critical balance conjecture results in a predicted power spectrum for ideal MHD given by the expressions,
\begin{equation}
	E(k_{\bot}) \propto k_{\bot}^{-5/3},
\end{equation}
\begin{equation}
	E(k_{\parallel}) \propto k_{\parallel}^{-2},
\end{equation}
as is discussed in \cite{Schekochihin2009}. These relations indicate that the spectrum along the field lines is softer than that across the field lines in ideal MHD turbulence. Indeed, our simulations produce this same anisotropy, but as will be discussed further in Section 5, our spectra are much softer due to physics separate from fully-developed MHD turbulence. We note that the critical balance conjecture (Equation 7) only formally holds in the regime where the mean magnetic field $B_0$ is much stronger than the fluctuating component $\delta B$ ($\delta B/B_0\ll1$), i.e. in the presence of a strong guide field which maintains the anisotropic development of turbulence. Because the ICM is a high $\beta$ plasma, we are probing the dynamo regime of MHD turbulence, i.e. $\delta B/B_0 \gtrsim 1$. Plasma dynamics within this regime are a topic of active research and we do not attempt to compute scaling laws for turbulence within this regime. Rather, we show that because the field in Simulation 1 remains relatively well-ordered throughout the simulation, we can formally define $k_{\parallel}$ and $k_{\bot}$, computing anisotropic power spectra. 

\section{The Model and Computational Setup}\label{simulations}

This work builds off RBS. Thus we set up the same simulation, with a plane-parallel Cartesian $(x,y,z)$ geometry in a 2$h$ $\times$ 2$h$ $\times$ 8$h$ domain, and a gravitational field defined by
\begin{equation}
	\textbf{g} = -\frac{g_0}{1+z/z_0} \hat{\textbf{z}},
\end{equation}
where $z_0$ is a trapping height for the g-modes which is taken to be 10$h$, larger than the height of the domain. Here $h = c^2_{s}/g_0$ is the scale height at the base of the ICM atmosphere and $c_{s}$ is the isothermal sound speed. RBS found that g-mode trapping by a sufficiently steep entropy gradient would serve to prevent the production of volume-filling turbulence. We choose a weakly trapped atmosphere in order to maximize the induced turbulent energy and fill our model ICM volume. The atmosphere is assumed to be initially isothermal, with a gas pressure $P$ and density related by $P = c^2_{s} \rho$  which satisfies the equation of hydrostatic equilibrium for a magnetized atmosphere,
\begin{equation}
	(1 + 1/\beta)\nabla P = \rho \textbf{g}.
\end{equation}
Here, $\beta$ is the plasma $\beta$ given by
\begin{equation}
	\beta = \frac{8 \pi P}{|\textbf{B}|^2},
\end{equation}
and is defined to be uniform in the initial atmosphere. This configuration yields an initial atmospheric profile described by
\begin{equation}
	\rho = \rho_0 \Bigg(1 + \frac{z}{z_0}\Bigg)^{-z_0/(1+1/\beta)h}.
\end{equation}
We then simulate jet impacts with the ICM as thermal ``bombs'' detonated at the base of the atmosphere. These ``bombs'' are implemented in the code by forming a radial over-pressurization at a height $z_{bub} = 0.2h$ in the atmosphere of size $r_{bub} = 0.1h$ and pressure $\Delta P_{bub} = 5\rho_0 c^2_{s}$. These bombs are detonated stochastically with varying $x$ and $y$ position with a mean recurrence time of $t_{inj}$ = 10 $h/c_{s}$ until a time $t_{stop}$ = 200 $h/c_{s}$. The simulation is then run out until a time 500 $h/c_{s}$. We focus on the late-time dynamics.

We use version 4.2 of the PLUTO code \citep{Mignone2007} to model the dynamics of our atmosphere under AGN driving. PLUTO evolves the equations of ideal magnetohydrodynamics in conservative form,
\begin{equation}
	\frac{\partial \rho}{\partial t} + \nabla \cdot (\rho \textbf{v}) = 0,
\end{equation}
\begin{equation}
	\frac{\partial}{\partial t} (\rho \textbf{v}) + \nabla \cdot \Big[\rho \textbf{v} \textbf{v} + \Big(P + \frac{|\textbf{B}|^2}{2} \Big) \mathcal{I} + \textbf{B} \textbf{B} \Big] = -\rho \nabla \Phi,
\end{equation}
\begin{equation}
	\frac{\partial}{\partial t} (E + \rho \Phi) + \nabla \cdot \Big[ \Big(E + P + \frac{|\textbf{B}|^2}{2} + \rho \Phi \Big) \textbf{v} \Big] = 0,
\end{equation}
where $\textbf{v}$ is the fluid velocity, $\mathcal{I}$ is the unit rank-two tensor, E is the total energy density of the fluid,
\begin{equation}
	E = u + \frac{1}{2} \rho |\textbf{v}|^2,
\end{equation}
and $\Phi$ is the gravitational potential which for our problem is, from Equation 10,
\begin{equation}
	\Phi = g_0 z_0 \ln \Bigg(1 + \frac{z}{z_0} \Bigg).
\end{equation}
We solve these equations on a 256 x 256 x 1024 uniformly-spaced Cartesian grid in order to properly resolve the beginnings of the turbulent cascade. The upper boundary is set with an outflow condition, and as will be discussed more completely in Section 5, we implement periodic boundary conditions in the horizontal dimensions so that Fourier transforms can be properly used to analyze the flow field. We use constrained transport \citep{Evans1988} to control the $\nabla \cdot \textbf{B}$ = 0 condition with an {\tt hllc} \citep{Toro1994} solver and second-order Runge-Kutta time-stepping. In order to track the evolution of the simulation, we employ two tracer fluids, denoted by $\mu_1$ and $\mu_2$. The tracer fluid $\mu_1$ is injected with the bubble and thus serves to trace the hot bubble plasma. The tracer $\mu_2$ is initially distributed in layers separated by 0.5$h$, providing a Lagrangian view of the fluid in order to track the production and development of g-modes and turbulence. Each simulation is run on 480 cores for approximately 100 hours on the University of Maryland's Deepthought2 supercomputer.

\section{Magnetic Field Morphology}\label{field_morphology}

We begin by discussing the simplest field configuration, a magnetic field purely in the $y$-direction with a strength of $\beta$ = 100 (hereafter, Simulation 1). In order to analyze the dynamics of this simulation, we will compare Simulation 1 to a hydrodynamic control simulation with identical initial conditions and identical driving. In reality, in order to employ exactly the same algorithms, this control simulation is not truly hydrodynamic. Rather, we use the same RK2 integrator with constrained transport $\nabla \cdot \textbf{B}$ control and implement an extremely weak field ($\beta$ = 10$^6$). We can determine that this simulation behaves identically to a hydrodynamic simulation (RBS) by taking power spectra both along and across the initial magnetic field direction. The power spectra match along and across the initial field configuration, implying that the anisotropy induced by the magnetic field is either dynamically unimportant or is so scrambled by the fluid motions that the field will do little to effect the overall evolution of turbulence.

\subsection{Magnetic Draping}\label{magnetic_draping}

Magnetic draping has two primary effects which allow us to evaluate its presence: 1) geometrically, the field lines should be lying across the bubble interface, stretched from their initial position due to being drawn up by the bubble, and 2) the strength of the field should increase, resulting in a drop in the plasma $\beta$ as field is built up at the bubble-ICM interface. Both of these trends are manifest in Figure 1. We note that the magnetic field increases to a $\beta$ of 8.72, and the magnetic energy density increases by nearly a factor of 10. Thus, even though our simulations begin with a weak magnetic field, the draping process strengthens the field to a level of dynamic importance. With this strength, the magnetic field can effectively stabilize against the parallel modes of the Rayleigh-Taylor instability. 

\begin{figure}
\centerline{
\hspace{0.3cm}
\psfig{figure=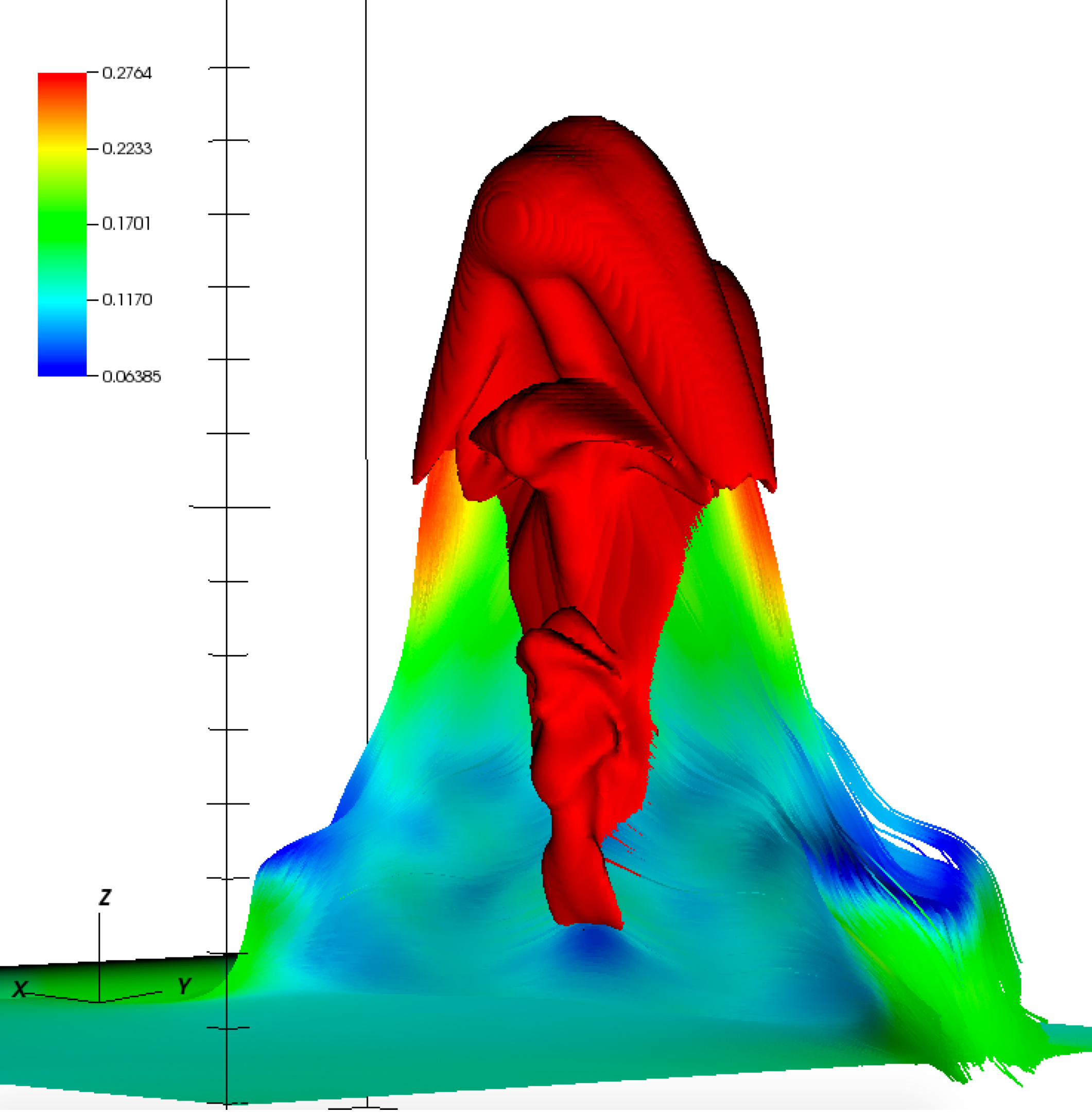,height=0.3\textheight} 
}
\caption{Visualization of bubble (red) with magnetic field lines in the $y$-direction draped across it at time 8 $h/c_s$. The field lines are colored by magnitude, with red colors representing the minimum $\beta$ value of 8.72. The crown of the bubble extends to a height of 1.6$h$ and the large axis tick represents a height of 1$h$.} 
\label{fig:draped_bubble}
\end{figure}

The magnetic field affects the morphology of the bubble in two separate ways. The bubbles do not maintain a spherical shape. Rather, they are compressed into sheets by the confining tension of the surrounding magnetic field. Physically, this can be accounted for by the field's natural restriction of perpendicular motion, forcing the path of least resistance to be upward instead of spherically outward as in a hydrodynamic simulation (see Figure 1 of RBS).

In addition, the field appears to drop in strength at the nodes of the field lines (shown in blue in Figure 1). This decrease in field strength is relatively symmetric about the bubble center and may be due to a decompression of the field in response to the stretching between the nodes. Because the field is anchored only to the surrounding fluid, an increase in field strength from the draping layer will result in a decreased tension about the bubble relative to the draping layer. 

\subsection{Suppression of Bubble Mixing}\label{mixing}

The earliest production of turbulence within the simulations comes not from the decay of g-modes, but from the shredding of the hot bubble plasma. In a hydrodynamic simulation, the shredding process allows efficient mixing of the bubble plasma, evolving the entropy of the bubble material to that of the surrounding plasma. In this way, hydrodynamic instabilities are a natural limit on the height of the bubble. As was stated in Section 1, magnetic fields provide a boundary between the two fluids of differing entropies, suppressing the instabilities which efficiently mix the bubble and ICM material.

The suppression of mixing appears to move beyond the initial shredding of the bubble. At late times, we see the Lagrangian tracer originally distributed in layers throughout the simulation, poorly mixed within the bubble debris layer (Figure 2). The density of the tracer fluid in the middle of the layers is approximately 50\% greater in Simulation 1 when compared to the hydrodynamic control run, indicating that magnetic fields suppress mixing. Our results are consistent with \cite{ZuHone2011} in that magnetic fields appear to suppress mixing between the hot bubble material and the surrounding ICM.

\begin{figure}
\centerline{
\hspace{0.3cm}
\psfig{figure=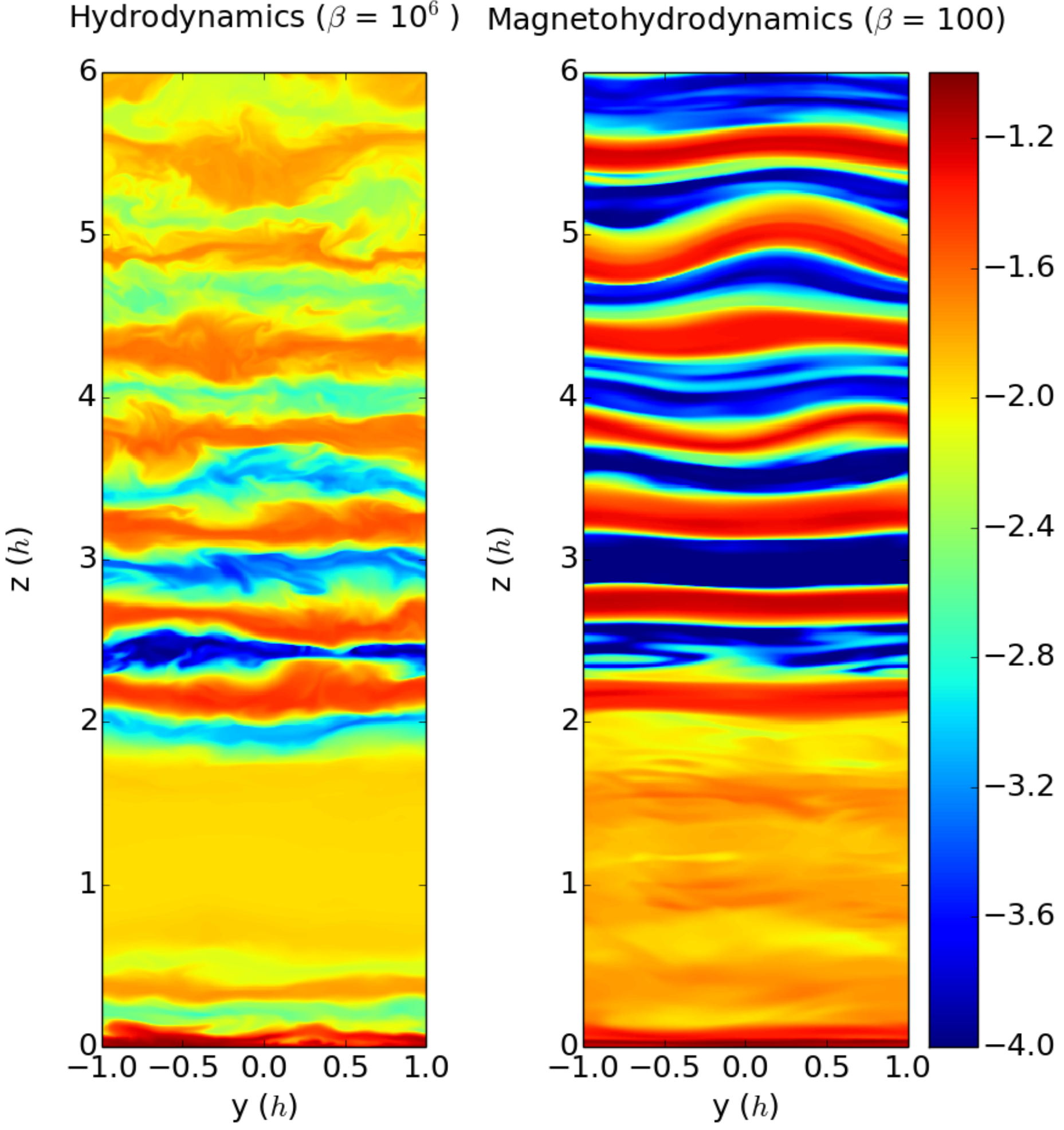,height=0.4\textheight} 
}
\caption{The passive tracer fluid (plotted with a logarithmic color scale) provides a Lagrangian view of the simulation domain. For the hydrodynamic ($\beta$ = 10$^6$) case, the fluid is well mixed and the domain exhibits volume-filling turbulence. With a $\beta$ of 100, turbulence is suppressed and the tracer remains minimally mixed.}
\label{fig:mixing}
\end{figure}

\subsection{The Effect of Field Geometry on Draping}\label{geometry}

The magnetic RTI is the primary cause of bubble break-up; however, the instability acts only across the field lines. In this section, we present a modified field geometry, a magnetic field defined by
\begin{equation}
	B_x = B_0 \cos(\kappa z), 
\end{equation}
\begin{equation}
	B_y = B_0 \sin(\kappa z),
\end{equation}
\begin{equation}
	B_z = 0,
\end{equation}
where $B_0$ is the magnetic field strength, defined with $\beta$ = 100. This field is a helical field geometry, i.e. the field twists with increasing height. \cite{Stone2007} argue that this field defined with a sufficiently high $\kappa$ value will suppress the magnetic RTI. The argument is that as the bubble rises, field from all directions will be encountered, and draping will proceed to form a ``mesh'' of field, suppressing the magnetic RTI in all directions. This geometry has previously been studied by \cite{Ruszkowski2007} who found that a helical field provided increased stability due to helicity conservation. These authors studied both the effects of external magnetic fields and internal fields within the bubble, finding that external fields tended to better stabilize bubbles when compared to internal fields. Motivated by these results, we choose an external helical field geometry in order to cause better preservation of the bubbles, suppression of instability-driven mixing between the bubble and ICM plasma, and amplification of g-modes which can then transition into turbulence.

For our simulations, we choose a $\kappa$ such that the field rotates once per scale height, i.e. 8 times total throughout the whole domain. We found that with noticeably tighter windings of the field, numerical reconnection became an issue, and the field effectively reconnects away, reducing results to that of a hydrodynamic simulation. This effect is non-physical since even a tightly-wound field is unable to reconnect in the absence of resistivity. Thus, we will only present results for the one twist per scale height case, i.e. $\kappa$ = $2 \pi h^{-1}$.

We found that draping proceeded similarly to Simulation 1 with one notable exception: the bubble was twisted by the helical field. When the bubble encountered field, the bubble proceeded to amplify the field until it became dynamically important to the bubble's evolution. The field then induced a pressure gradient about the bubble axis, causing the bubble to rotate about the bubble center. Eventually, with increasing height in the atmosphere, the bubble expanded and became subject to the magnetic RTI, allowing the bubble to be shredded in long, twisted ribbons (Figure 3). In order to determine the role of the bubble pressure on exciting this instability, we performed a run where a bubble began in pressure equilibrium with the surrounding atmosphere. This bubble did not rise as high in the atmosphere as the over-pressurized bubble before being shredded by the magnetic RTI. In addition, the bubble does not expand as significantly without the added pressure, resulting in less displacement of fluid on large scales, and thus weaker g-modes. Comparing the ambient incompressible energy between the over-pressurized and pressure equilibrium bubbles, we find approximately 27\% more energy in g-modes and turbulence for the over-pressurized bubble simulation compared to the pressure equilibrium run (see Section 5.1 for discussion of compressible vs incompressible modes).

This ribbon shape is likely unphysical, being a consequence of our chosen geometry. Our simulations are not attempting to reproduce the exact morphology of the bubbles, but rather produce sufficient g-modes to account for the turbulent energy density measured in the ICM. As will be discussed further in Section 5, we are unable to noticeably amplify g-modes through the use of magnetic draping. In subsequent sections, we will discuss the cause of this inefficiency and a secondary effect which acts to suppress turbulence: magnetic tension. 

\begin{figure}
\centerline{
\hspace{0.3cm}
\psfig{figure=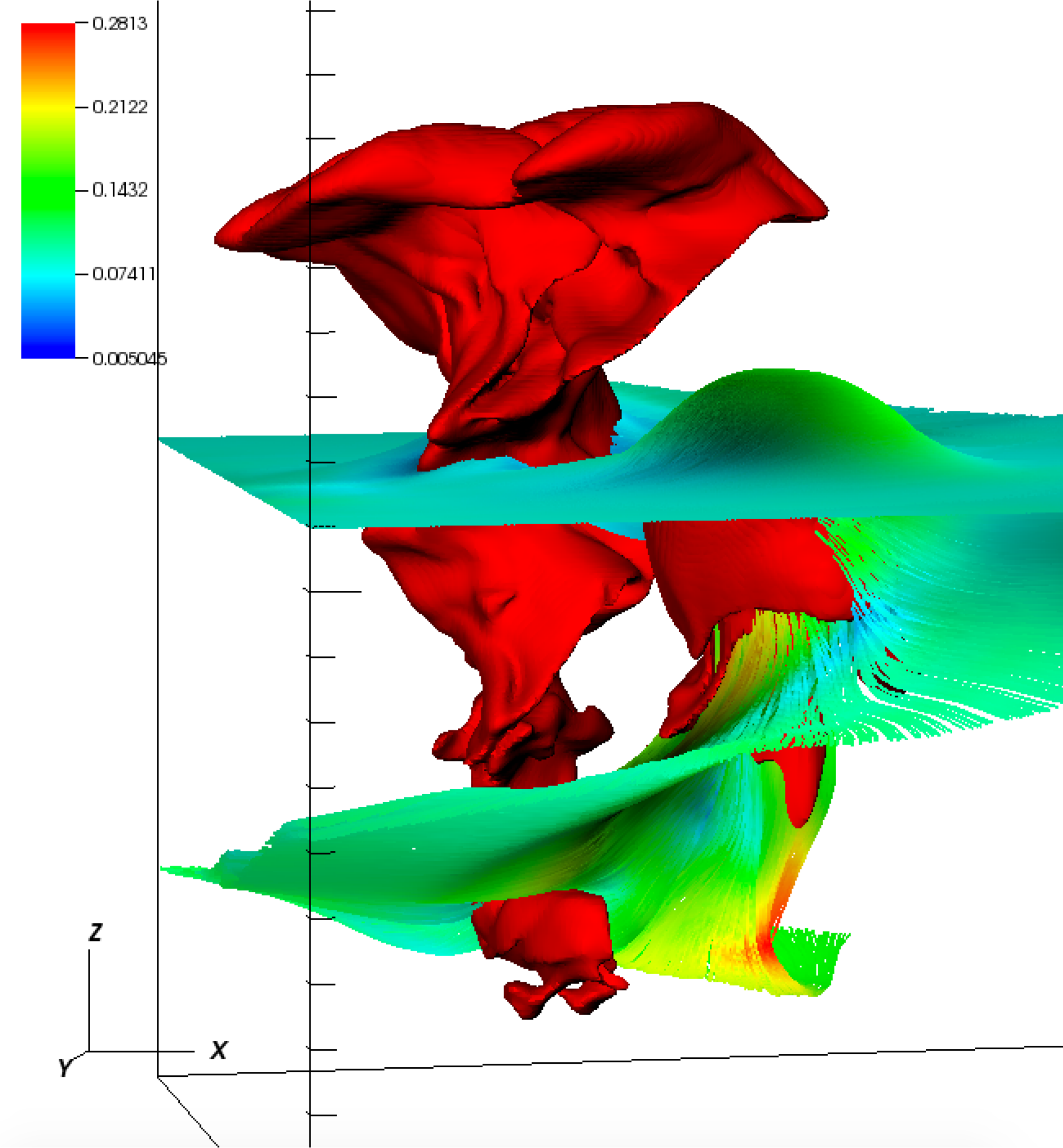,height=0.35\textheight}
}
\caption{Visualization of bubbles (red) with magnetic field lines from helical field draped across them at time 10 $h/c_s$. The field lines are colored by magnitude, with red colors representing the maximum field. The field begins draping over the spherical bubble as the bubble twists through the field as demonstrated by the smaller foreground bubble. Once the bubble has expanded sufficiently, the bubble-ICM interface is magnetic Rayleigh-Taylor unstable and shreds into ribbons as demonstrated by the larger bubble. The crown of the larger bubble extends to a height of $\sim$ 1.75$h$ and the large axis tick represents a height of 1$h$.}
\label{fig:helical_bubble}
\end{figure}

\section{Suppression of Turbulence}\label{turbulence_suppression}

In the subsequent sections, we will demonstrate that 1) AGN-driven turbulence within our model ICM is inefficient, with magnetic fields suppressing rather than amplifying turbulence. 2) Our driving mechanism efficiently produces magnetosonic waves; however, incompressible modes such as g-modes and turbulence are primarily lost to the magnetic field and dissipation in the bubble debris region. 3) Magnetic tension suppresses the nonlinear decay of g-modes to turbulence, and the magnetic field geometry is critical to this suppression. 4) The field geometry modifies the trapping of g-modes, and though g-modes collide within our atmosphere, the interaction remains quasi-linear for the helical field simulation, forming a beating frequency between g-modes.

\subsection{Compressible and Incompressible Energy Evolution: Evaluating the Efficiency of AGN-Driven Turbulence}\label{comp_v_incomp}

Bulk turbulence and g-modes in the ICM are manifest in incompressible motions of the fluid. Thus, we can study the evolution of both of these features by decomposing the velocity into compressible and incompressible motions and removing the contribution of the compressible modes (magnetosonic waves). This separation of velocity modes can be accomplished through the use of a Helmholtz decomposition.

For any velocity field $\textbf{v}$, the field can be written
\begin{equation}
	\mathbf{v} = \mathbf{v}_c + \mathbf{v}_i = \nabla \psi + \nabla \times \mathbf{a},
\end{equation}
where $\psi$ is a scalar potential and $\mathbf{a}$ is a vector potential. The compressible velocity field potential $\psi$ can be solved for through a Poisson's equation for the compressible modes,
\begin{equation}
	\nabla^2 \psi = \nabla \cdot \mathbf{v}.
\end{equation}
From here, $\textbf{v}_c$ can be solved for using standard Fourier techniques. Note that our use of periodic boundary conditions guarantees the applicability of the fast Fourier transform (FFT) toward analyzing our velocity field. Therefore we are able to decompose our velocity field into compressible and incompressible modes.

Figure 4 displays the results of this calculation. By computing the total energy present in the incompressible modes at the end time of 500 $h/c_{s}$, i.e. once initial excitations of turbulence from the bubble shredding have calmed down and the remaining kinetic energy is due to g-mode turbulence, we can calculate an efficiency for AGN-driven g-mode turbulence. 

Though we attempted to control for the driving in our simulations by using identical initial conditions, our driving mechanism was tied to a series of calls to a random number generator and hence to the integration time-step in the simulation. As such, identical initial conditions yielded thermal bombs detonating at different times depending on the magnetic field strength and geometry. The helical field simulations were driven by a total of 25 thermal bombs while the hydrodynamic control run was only subject to 27 and the unidirectional field 28. We divide the remaining ambient incompressible energy by the total injected energy to compute the efficiency of driving for the helical field simulation:
\[
	E_{inj} = \frac{3}{2} \Delta P V \times N_{bombs}.
\]
\[
	\mathrm{Efficiency} = E_{end}/E_{inj} = \frac{1.13 \times 10^{-4}}{0.79} \approx 0.014 \%.
\]
We find an efficiency of 0.014\% for the helical field simulation, more than a factor of 3 \textit{less} than the hydrodynamic simulation (0.046\%). The unidirectional field case (Simulation 1) shows similar inefficiency (0.023\%). Indeed, rather than increase the efficiency of driving turbulence, magnetic fields seem to \textit{suppress} turbulence.

The physical reasoning behind this effect appears to be two-fold: 1) incompressible motions rapidly dissipate in the bubble debris region during driving and 2) the magnetic tension force prevents the nonlinear ``breaking'' of g-modes. Each of these effects will be evaluated and discussed in the following subsections.

\begin{figure*}
\hbox{
\psfig{figure=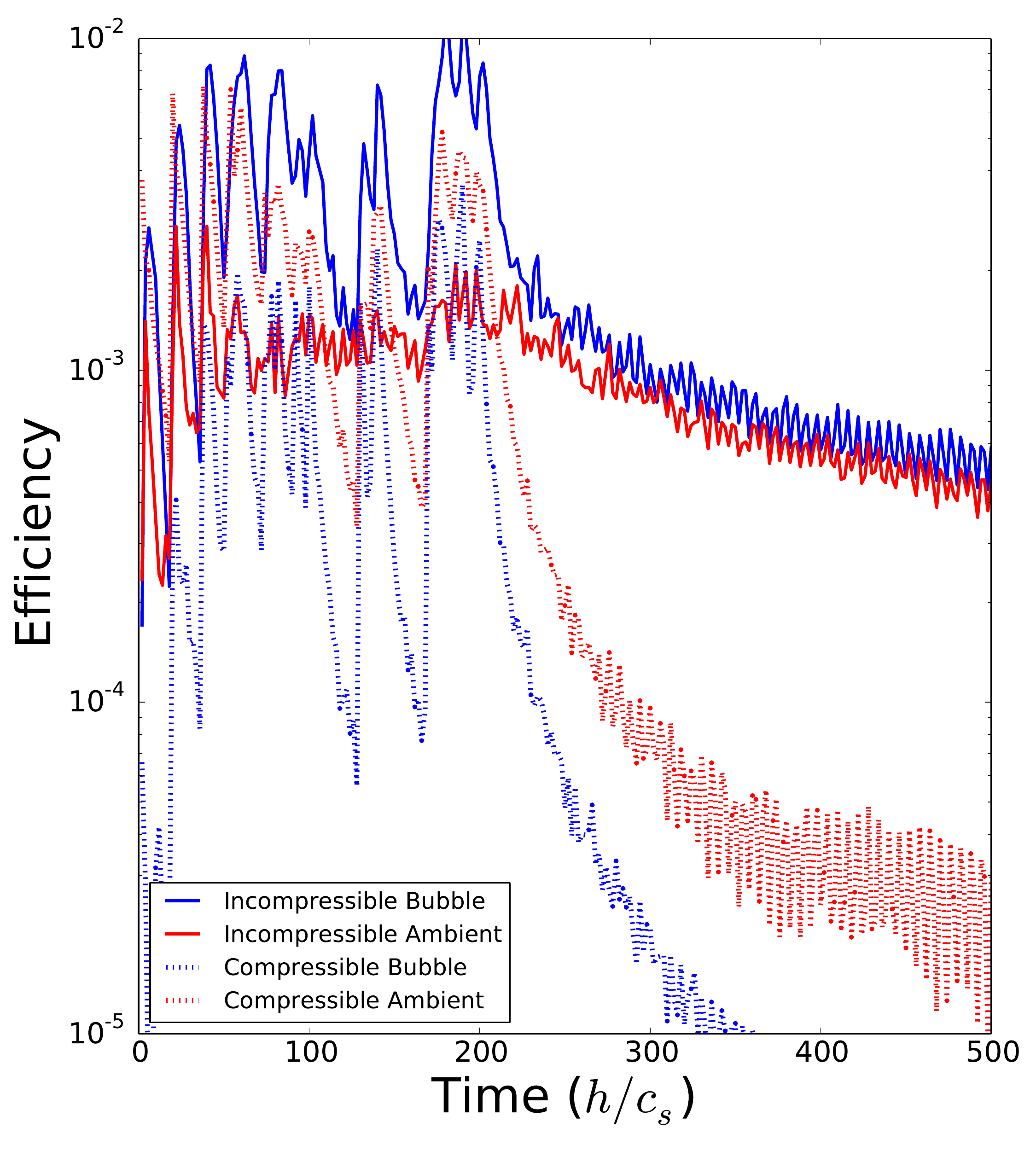,width=0.50\textwidth}
\psfig{figure=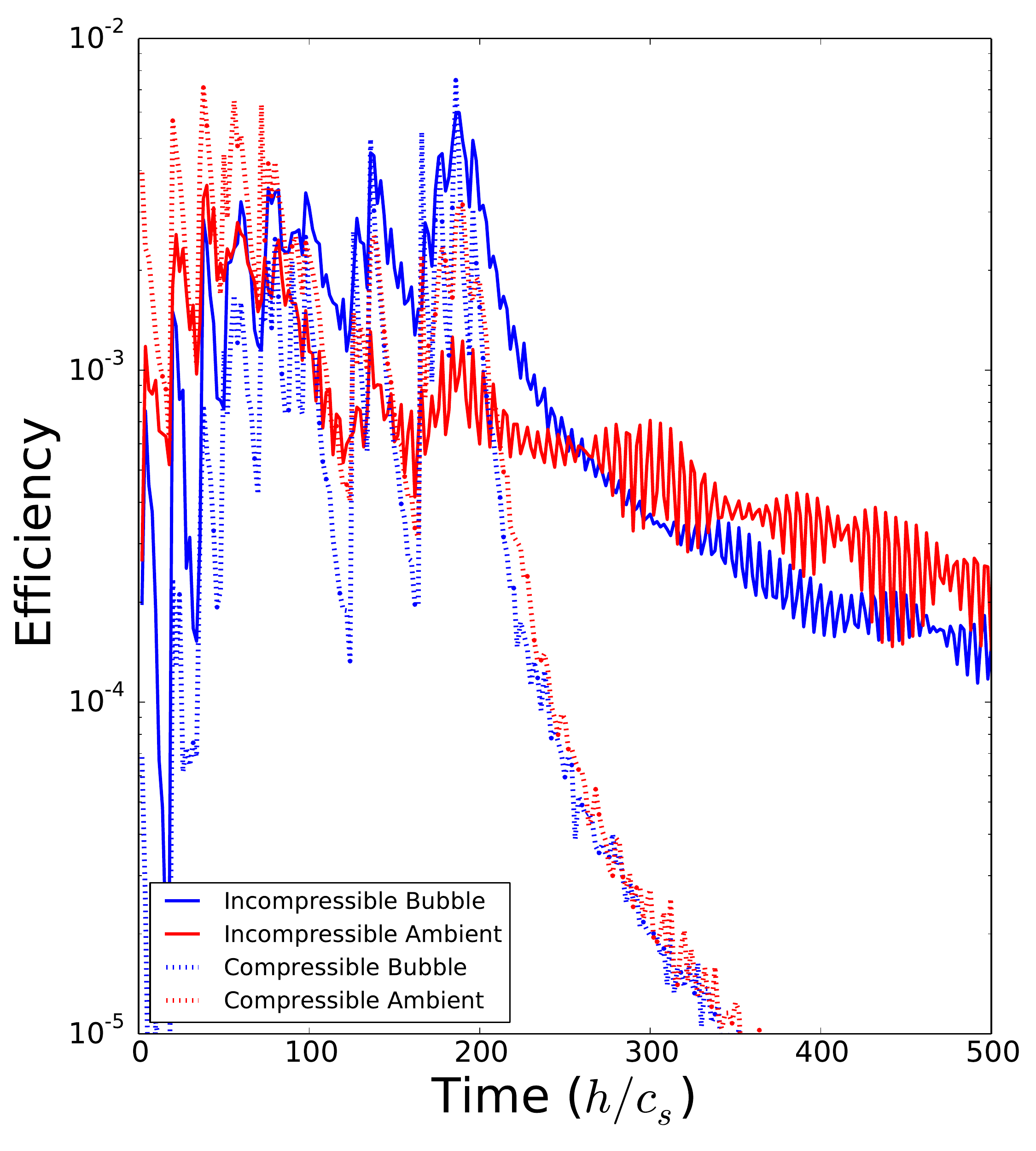,width=0.50\textwidth}
}
\caption{Volume-integrated kinetic energies in compressible and incompressible modes for the hydrodynamic control (left) and the helical field simulation (right). The incompressible and compressible kinetic energies display the relative contribution of g-modes and turbulence (incompressible modes) vs that of magnetosonic waves (compressible modes) as a function of time in our simulation domain. We find that 0.014\% of injected energy ends up in incompressible modes by $t$ = 500 $h/c_s$ for the helical field simulation, more than a factor 3 less than that measured in the hydrodynamic control (0.046\%). Note the high frequency oscillations apparent in both simulations correspond to the frequency of g-modes, while the lower frequency envelope seen only in the helical field simulation likely corresponds to a beating frequency produced by the quasi-linear interference of two g-modes (see Section 5.6).
}
\label{fig:velocity_decomposition}
\end{figure*}

\subsection{The Role of Magnetosonic Waves}

We have argued that the driving of turbulence is inefficient. Yet, if energy from our thermal bombs is not transferred to incompressible g-modes, what happens to the energy? The majority of kinetic energy is transferred to compressible modes, i.e. magnetosonic waves, which leave the simulation domain during the driving phase ($t  < $ 200 $h/c_s$). If we assume the peak total compressible energy from the first explosion to be an accurate measure of the kinetic energy from each thermal bomb, we find energies of 0.088 and 0.079 $\rho_0 c_s^2 h^3$ for the unidirectional and helical field simulations respectively, corresponding to 10\% of the injected energy for both simulations, approximately equal to the 10\% efficiency of the sound waves in the hydrodynamic simulation. 

The efficiency of driving magnetosonic waves is likely a consequence of our driving mechanism. The instantaneous over-pressurizations which we use to inject energy into our simulations are an oversimplification of the complex jet-ICM interaction. However, the observations of cavities and weak shocks \citep{Jones2002, Million2010, Randall2015} in the ICM support a supersonic model for bubble inflation \citep{Fabian2006}. Models of observed ICM shocks support the argument that bubbles should be inflated slowly enough such that only $\sim$ 20\% of the injected energy is carried by shocks \citep{Forman2017, Zhuravleva2016, Zhuravleva2017}. In addition, idealized simulations of the bubble inflation process argue that with constant energy injection, the bubble will initially expand supersonically before gradually becoming subsonic \citep{Tang2017}. In the instantaneous injection limit (relevant for our simulations), a maximum of $\approx$ 12\% of injected energy is carried away by sound waves. Thus, our 10\% estimates of the total energy in compressible modes are consistent with both observations and theory despite our use of ``thermal bombs'' to inject energy.


\subsection{Energetics of a Driven Stably-Stratified Magnetized Atmosphere}

To understand the energy budget available to drive g-modes, we must understand how the injected energy is distributed between the available channels, namely kinetic energy, thermal energy, gravitational potential energy, and magnetic energy. By observing a single detonation of a thermal bomb, we can analyze the energy distribution between these 4 channels (Figure 5). 

In Figure 5, we compare the overall energetics between the unidirectional field simulation (Simulation 1), helical field simulation (Simulation 2), and the hydrodynamic control run. In Simulations 1 and 2, energy is partitioned into stretching the magnetic field. Tension from magnetic draping slows the rise of the bubble compared to the hydrodynamic run. The atmosphere is unable to expand efficiently, and the gravitational energy thus does not increase as quickly while the thermal energy remains higher compared to the hydrodynamic run. The kinetic energies of the magnetized runs drop more rapidly than that of the control run; kinetic energy is rapidly dissipated in the bubble debris region. 

Indeed, even if we take extremely conservative estimates of the energy available for driving turbulence by computing the incompressible kinetic energy immediately after driving at $t$ = 200 $h/c_s$, we find efficiencies of 0.08\% and 0.06\% for Simulations 1 and 2 respectively, still orders of magnitude too low to account for the Hitomi results. Comparing the peak energies between thermal energy and kinetic energy in the magnetized runs, we find approximately 18\% of the injected energy is driven into kinetic energy. Compressible modes account for 10\% of the total energy, leaving $\sim$ 8\% to be rapidly dissipated in the bubble debris region and 2 orders of magnitude less energy available to drive g-modes and turbulence.

\begin{figure}
\centerline{
\hspace{0.3cm}
\psfig{figure=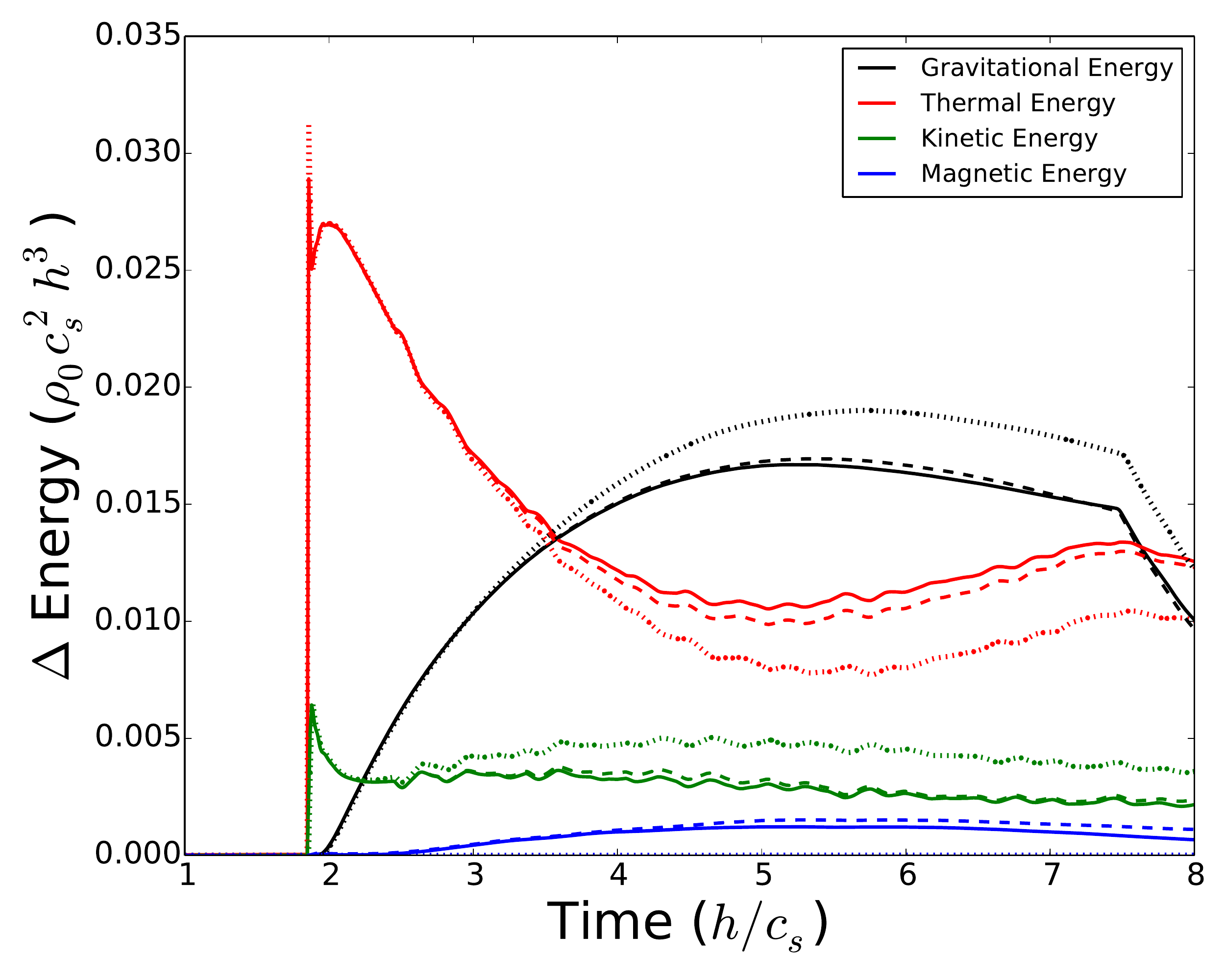,height=0.28\textheight}
}
\caption{Change in energy from the detonation of a single thermal bomb. The helical field simulation is shown by the bold line, the unidirectional field the dashed line, and the hydrodynamic control the dotted line. During the driving phase, energy is lost to the stretching of the magnetic field and dissipation in the bubble debris region, leading to a decrease of available kinetic energy. }
\label{fig:single_explosion}
\end{figure}

\subsection{Evaluating the Efficiency of G-Mode Production}

\begin{figure*}
\hbox{
\psfig{figure=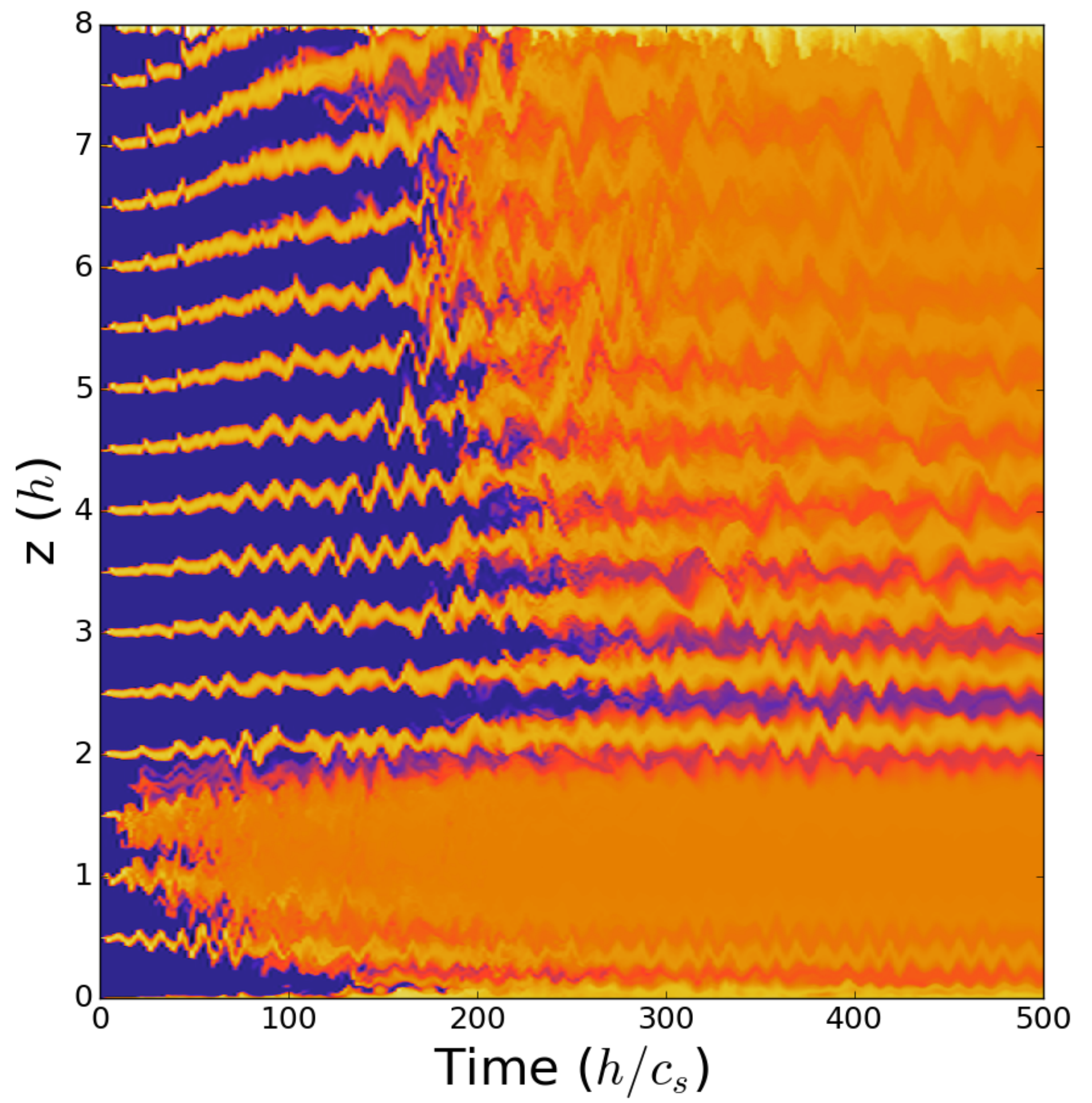,width=0.50\textwidth}
\psfig{figure=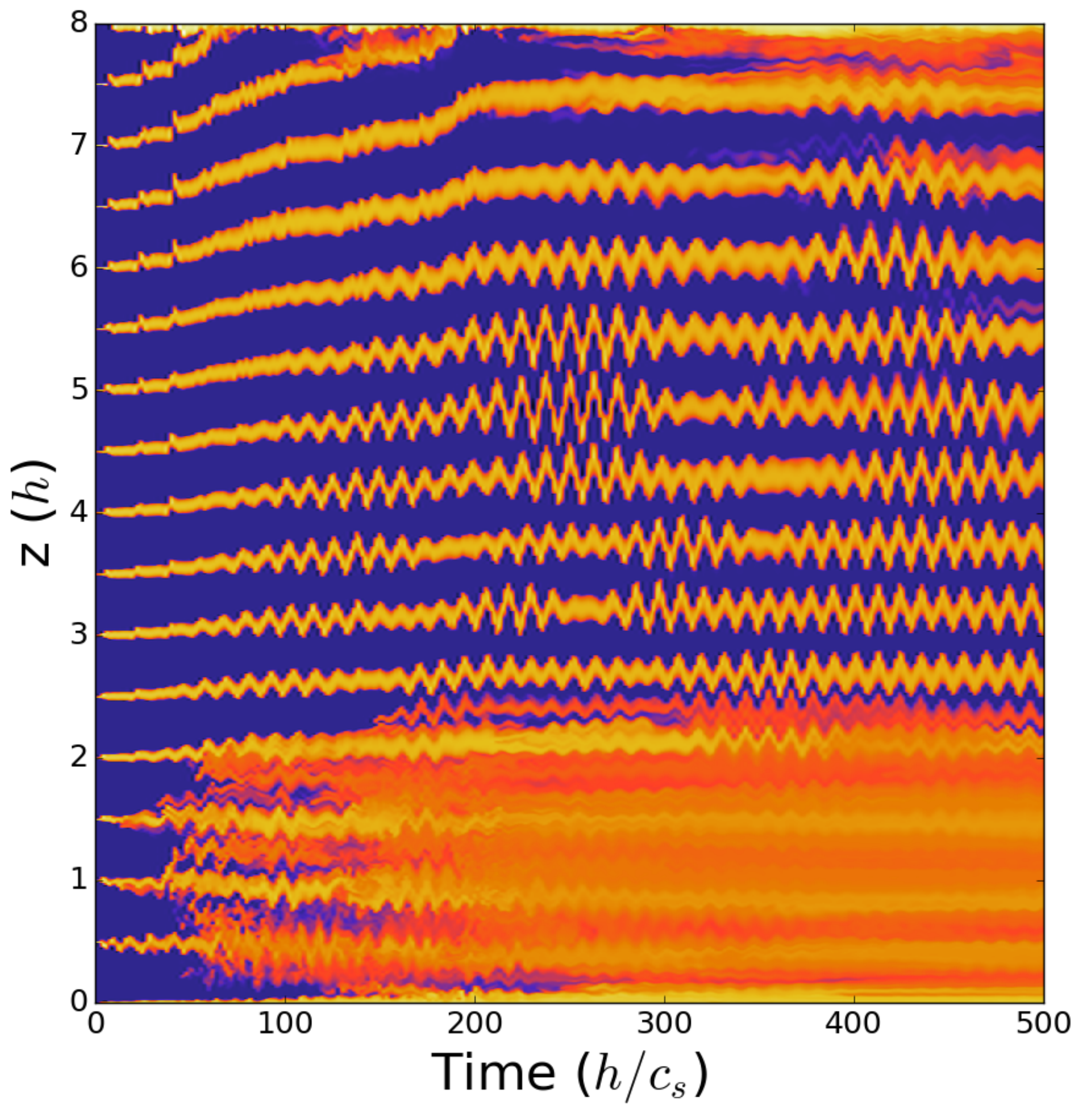,width=0.50\textwidth}
}
\caption{Space-time evolution of core sample of the Lagrangian tracer $\mu_2$ tracing the evolution of g-modes in the unidirectional field simulation (left) and the helical field simulation (right). The fluid layers remain distinct in the helical field simulation, indicating that the g-modes never break and transition into turbulence. While there may appear to be larger amplitude g-modes in the helical field simulation immediately after driving at 200 $h/c_s$, the difference is likely negligible, indicating that magnetic draping does not significantly amplify g-modes in our simulations.}
\label{fig:space_time_gmodes}
\end{figure*}

In order to determine the efficiency of g-mode production, we analyze the tracer fluid $\mu_2$. This fluid, originally distributed in layers with a separation of 0.5$h$ throughout the domain, provides a Lagrangian perspective on the fluid motions. The ``strength'' of g-modes is proportional to the wave's amplitude. Thus, the larger the g-mode amplitude, the more energetic the mode. We take a core sample (x = y = 0$h$) of the tracer $\mu_2$ and plot the evolution of this tracer as a function of time (Figure 6) for both the unidirectional and helical field simulations.

Figure 6 displays three notable features: 1) the amplitude of the largest g-modes in the helical field simulation are comparable to those of the unidirectional field simulations and hydrodynamic simulations (not shown), 2) the g-modes never break and transition into turbulence in the helical field case, while there is clear breaking in the unidirectional field simulation, and 3) the g-modes in the helical field simulation seem to decay and then rise up again later on in the simulations. 

Even though the bubbles appear to maintain more coherent structures in the helical field simulations, draping of a helical field does not save the bubbles from shredding due to the magnetic RTI. Rather, the helical field modifies the instability to produce ribbons of bubble material weaving through the field lines. These ribbon structures, though they may appear to be more coherent than the vortex rings formed in the hydrodynamic simulations or the sheets formed in the unidirectional field simulations, may not actually displace much more material. Thus, magnetic draping does not appear to increase the strength of g-mode driving.

The suppression of the nonlinear decay of g-modes to turbulence is a consequence of the magnetic field geometry. When the field is in only one direction, the g-modes can still decay to turbulence while the helical field fully suppresses this nonlinear interaction. 

We note that because the unidirectional and helical field simulations display similar g-mode amplitudes, the lack of nonlinear interaction is in fact a consequence of the field geometry rather than simply the driving strength. For this reason (and based on the analysis presented in the next section), we conclude that the magnetic tension force is responsible for the suppression of the nonlinear interaction and ensuing g-mode turbulence. We do note however that there may be limits to this suppression, i.e. with much stronger driving, the g-modes might break and decay to turbulence.

\subsection{Power Spectra: Suppression of Turbulence}

The role of magnetic tension and the subsequent suppression of turbulence can best be evaluated through the use of power spectra. We can determine whether or not sufficient energy to offset radiative cooling reaches small scales by analyzing the shape of power spectra taken in Simulations 1 and 2.

First, we will perform a similar analysis to RBS and compute the power spectrum of the horizontal incompressible modes in each simulation. We compute power spectra in 1 grid cell slices and average over a slab 0.5$h$ thick, centered at a height of 5$h$. We then average the spectra over the times 300 to 400 $h/c_s$, allowing initial disturbances to settle and g-modes to decay to form a turbulent cascade. The choice of a height of 5$h$ eliminates the problem of anomalous shock heating due to sound wave steepening in our Cartesian atmosphere. This choice also implies that our spectra are taken in the ambient medium and all turbulence is due to the breaking of g-modes rather than the shredding of bubbles.
We define our horizontal power spectra with the expression,
\begin{equation}
	E(k_h) = \pi k_h \Big[|\widetilde{v}_{ix}(k_h)|^2 + |\widetilde{v}_{iy}(k_h)|^2 \Big],
\end{equation}
where $k_h$ is the magnitude of the horizontal wavenumber vector,
\begin{equation}
	\textbf{k}_h = k_x \hat{\textbf{x}} + k_y \hat{\textbf{y}},
\end{equation}
and the tilde denotes the Fourier transform of the vector field. Figure 7 displays the results of this calculation. We only analyze the $x$ and $y$ motions in the simulations. This choice is motivated by the expectation of stratified turbulence \citep{Zhuravleva2014} and because any $z$-vorticity (and thus fully consistent turbulence) would be due to motions in the $x$-$y$ plane. 

\begin{figure}
\centerline{
\hspace{0.3cm}
\psfig{figure=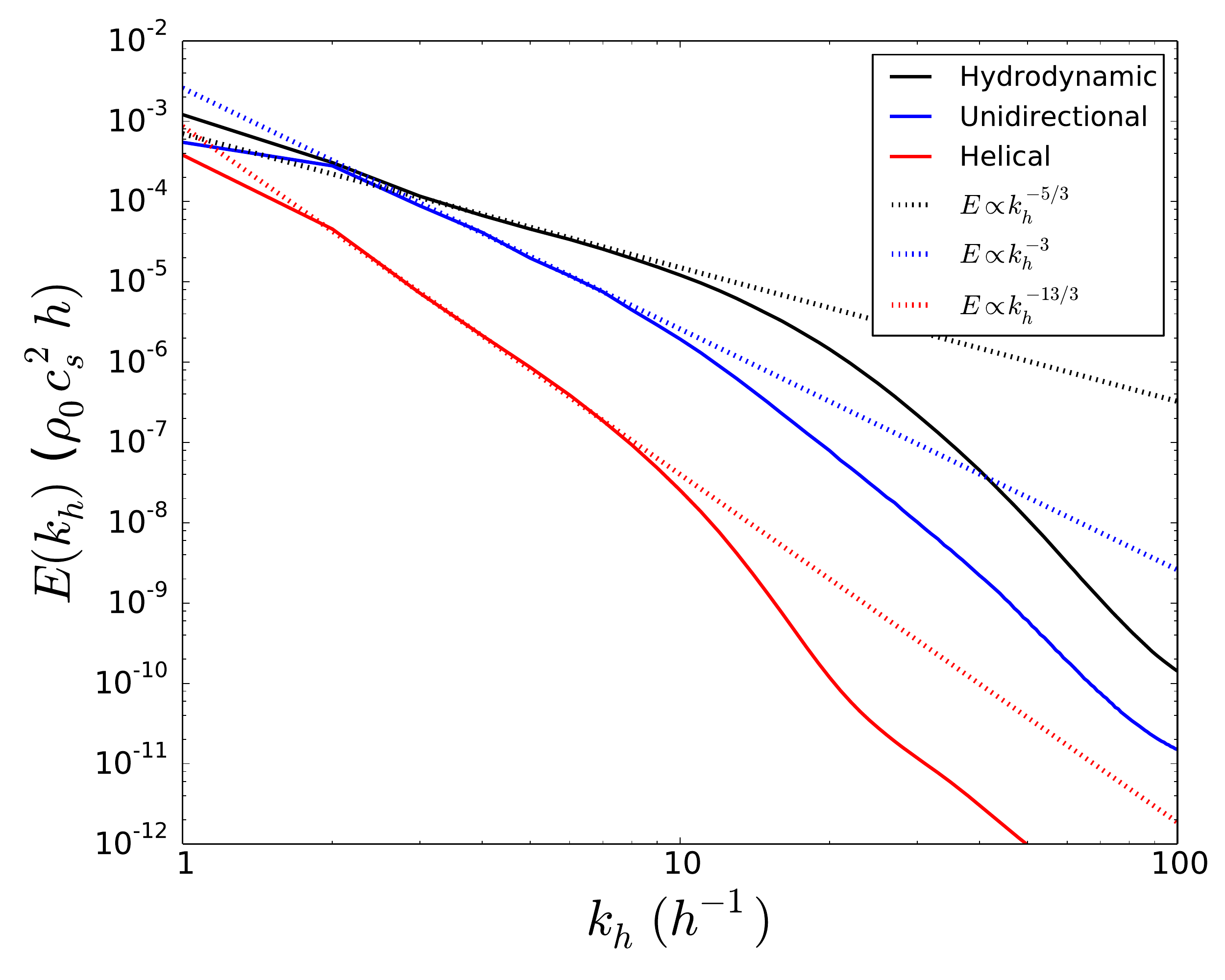,height=0.28\textheight}
}
\caption{The power spectrum of the unidirectional field simulation (blue) is softer than that of the hydrodynamic control run (black), an averaging effect due to the anisotropy imposed by the magnetic field. The power spectrum of the helical field simulation (red) is significantly softer than those of the other simulations, indicating that the direction of the magnetic field has a significant effect on the development of turbulence. The spectra measured in the helical field simulation indicate a separation of scales, with g-modes at the largest scales and weak breaking and dissipation at the smallest scales. Dotted lines show approximate fits to the measured slopes. Spectra are weighted by the number of thermal bombs driving the hydrodynamic control run (27).}
\label{fig:horizontal_power_spectrum}
\end{figure}

The horizontal power spectra show notable features. In the hydrodynamic control run, we see a clear Kolmogorov ($k_h^{-5/3}$) spectrum throughout the inertial range (1 $\leq k_h \leq$ 10 $h^{-1}$). Scales below $k_h$ = 10$h^{-1}$ correspond to $\sim$ 12 computational zones. Thus, we see a strong softening of the spectra below this scale, associated well with grid-scale dissipation.

For the unidirectional field simulation (Simulation 1), we find an approximate $k_h^{-3}$ spectrum, significantly softer than the Kolmogorov spectrum expected for self-consistent MHD turbulence. As we will show, this spectrum can be explained by the anisotropic development of turbulence. Along the field lines, the spectra are extremely soft due to the effects of magnetic tension while across the field lines, the spectra are nearly Kolmogorov. The $k_h^{-3}$ spectrum is thus an averaging effect which comes from measuring spectra in directions with components both along and across the magnetic field.

By time motions reach the dissipation scale in Simulation 1, we find a power of $\sim$ 1.7x10$^{-6}$ $\rho_0 c_s^2 h$, a factor of more than 6 less than the power measured in the hydrodynamic control (1.1x10$^{-5}$ $\rho_0 c_s^2 h$), indicating very little power at small scales. As expected, the helical field simulation (Simulation 2) shows the greatest suppression of turbulence. At the dissipation scale, we measure a power of only 2.4x10$^{-8}$ $\rho_0 c_s^2 h$, a factor of nearly 500 less than the power measured in the hydrodynamic control run. The inertial range spectrum for the helical field simulation shows an approximate $k_h^{-13/3}$ scaling which is not predicted within the standard critical balance theory of MHD turbulence. Thus, this spectral slope points more directly toward the physics of g-mode breaking, namely that energy remains in large-scale g-modes which slowly dissipate at small scales. 

In the previous section, we argued that turbulence was suppressed by the magnetic tension force which prevented the nonlinear decay of g-modes to turbulence. We can diagnose the role of magnetic tension by computing anisotropic power spectra along both the $k_{\parallel}$ and $k_{\bot}$ directions, where parallel ($\parallel$) and perpendicular  ($\bot$) are in reference to the magnetic field direction. The parallel and perpendicular spectra respectively are described by the following expressions,
\begin{equation}
	\begin{aligned}
		E(k_{\parallel}) = \pi k_{\parallel} \Big[|\widetilde{v}_{ix}(k_{\parallel})|^2 + |\widetilde{v}_{iy}(k_{\parallel})|^2 \Big], \\[1 pt]
		E(k_{\bot}) = \pi k_{\bot} \Big[|\widetilde{v}_{ix}(k_{\bot})|^2 + |\widetilde{v}_{iy}(k_{\bot})|^2 \Big],
	\end{aligned}
\end{equation}
where
\begin{equation}
	\begin{aligned}
		\textbf{k}_{\parallel} = \frac{\textbf{k} \cdot \textbf{B}}{|\textbf{B}|^2} \textbf{B}, \\[1 pt]
		\textbf{k}_{\bot} = \textbf{k} - \textbf{k}_{\parallel}. 
	\end{aligned}
\end{equation}

Anisotropic power spectra can only be computed in cases where the magnetic field direction is well-defined, i.e. when $\delta B/B_0 \ll 1$. Because our simulations are probing the dynamo regime, the field direction is subject to change throughout the simulation due to the motions of the fluid. Thus, the field direction is not formally well defined; however, we find that in the unidirectional field case, the field remains well-ordered enough over the timescales of our simulation to allow for the computation of valid power spectra both along and across the local field direction. The same well-ordered field is not present in the helical field simulation (see Figure 8). In the helical field case, horizontal slices include field with equilibrium position in that slice as well as field pushed into the slice by the g-mode oscillation. Therefore, a single horizontal slice can have field from completely different directions present, and there is no well-defined field direction.

\begin{figure}
\centerline{
\hspace{0.3cm}
\psfig{figure=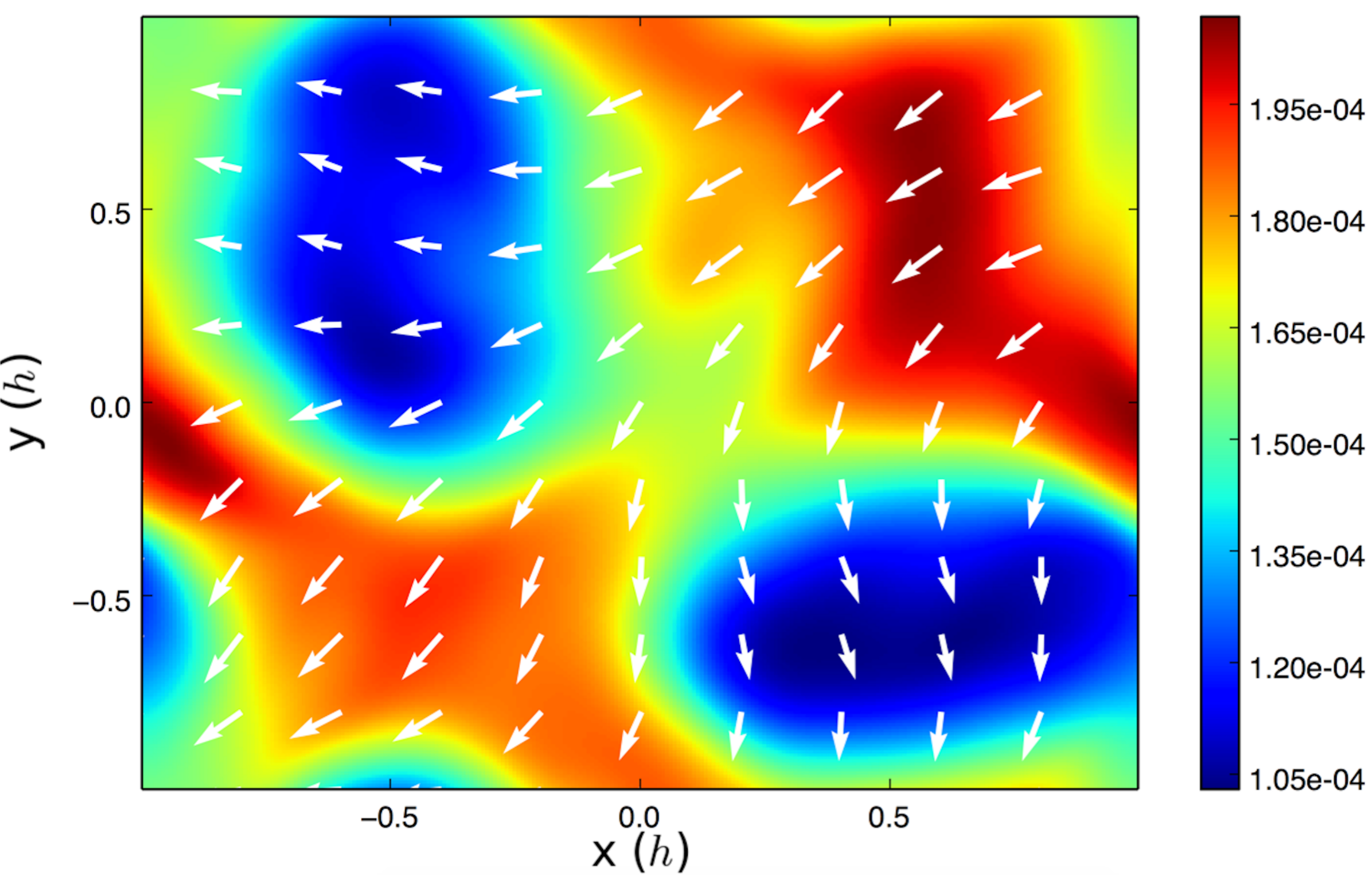,height=0.24\textheight}
}
\caption{Visualization of the magnetic field in an $x$-$y$ slice taken at height 5$h$ at time 308 $h/c_s$. The vectors are composed of the $x$ and $y$ components of the magnetic field, and the colors correspond to the magnetic energy density. We see that higher field regions and lower field regions correspond to fields with different equilibrium heights brought together through the oscillation of the g-modes. Because these fields have different directions, we are unable to properly define a field line direction to compute anisotropic power spectra in the helical field simulation.}
\label{fig:field_visualization}
\end{figure}

With the limitations on the helical field simulation in mind, we compute the anisotropic power spectra for the unidirectional field simulation and compare these spectra to the hydrodynamic control run in Figure 9. We note that the hydrodynamic spectra are really spectra along and across the \textit{initial} field direction. These spectra both show Kolmogorov-like slopes (as expected) with negligible differences in the normalization between them. Thus, we present the average of the parallel and perpendicular spectra for the hydrodynamic case in Figure 9. 

For the unidirectional field simulation, the spectrum across the field is approximately Kolmogorov within the inertial range with increased power in the inertial range when compared to the hydrodynamic run. Along the field lines however, the spectra are extremely soft, with a steep decay reminiscent of the scaling within the dissipation subrange. These spectra allow us to construct a description for the suppression of turbulence in our simulations and broadly explain why AGN-driven ideal MHD turbulence is so inefficient. 

Motions across the field lines are unaffected by the tension of the field. Rather, perpendicular motions passively advect the field lines and form self-consistent MHD turbulence, reminiscent of a Kolmogorov spectrum. Along the field lines, magnetic tension acts to both halt the nonlinear transition to turbulence as well as suppress the cascade of turbulent energy to small scales. Turbulent energy along the field lines is siphoned off into stretching the field, and the turbulence tends toward an anisotropic development, placing nearly all power in perpendicular modes before reaching the dissipation scale. The horizontal spectrum of the unidirectional field simulation is then well-described by an averaging effect between parallel and perpendicular power spectra. With a helical field, tension is able to act in all directions. Thus, magnetic tension halts the nonlinear interactions of the g-modes at all scales, causing bulk motions to remain at large scales throughout late times in the simulation. Kinetic energy is gradually transferred to thermal energy through dissipation as large-scale motions are brought to the dissipation scale through stretching the field. 

Figure 9 demonstrates that the nonlinear interaction time $(v_h k_h)^{-1}$ is long compared to that of the hydrodynamic control run, indicating that the nonlinear time may be too long to cause significant interactions on the buoyancy timescale. Just as magnetic tension suppressed the growth of the magnetic RTI and KHI at the bubble-ICM interface, this tension suppresses the nonlinear interactions which cause g-modes to decay into turbulence.

We note that these results are consistent with studies performed in the solar wind physics community, namely that magnetic fields can allow for the preservation of discrete modes (such as g-modes) for many nonlinear times even in the presence of broadband turbulence \citep{Ghosh2009}. Further work is required to determine the precise nonlinear mechanism which cause g-modes to decay into turbulence and the role of magnetic fields in inhibiting this mechanism; however, these questions lie beyond the scope of this work.

Finally, we estimate the critical $\beta$ at which this suppression process takes place. Since the hydrodynamic case ($\beta$ = 10$^6$) allows for g-modes to break uninhibited by the magnetic field and the $\beta$ = 100 fields suppress the nonlinear interactions which allow g-modes to decay, we presume that the critical $\beta$ for suppression lies between these values. The magnetic field becomes dynamically relevant at the point where the magnetic energy density is in an approximate equipartition with that of the horizontal velocity field, i.e.
\begin{equation}
	\frac{1}{2} \rho |\textbf{v}_h|^2 \sim \frac{|\textbf{B}|^2}{8 \pi}.
\end{equation}
This statement is equivalent to stating that $v_h$ $\sim$ $v_A$, where $v_A$ is the Alfv\`en velocity. To determine the value of $v_h$, we compute the horizontal velocity typical of g-modes decaying uninhibited by magnetic fields, i.e. the velocities from the hydrodynamic control run. The plasma $\beta$ can be rephrased in terms of the sound speed and Alfv\`en velocity, i.e. $\beta$ = 2 $c_s^2$/$v_A^2$. Thus, by setting the horizontal velocity equal to the Alfv\`en velocity, and computing $\beta$ at each grid cell in the ambient region (4-6 $h$), we find an estimate of $\beta_c$, the critical $\beta$ for turbulence suppression. An average over the times 200 - 500 $h/c_s$, i.e. after driving, yields $\beta_c$ $\sim$ 2000, a magnetic energy density more than an order of magnitude less than those implemented in Simulations 1 and 2. Therefore, within the ideal MHD framework, even weak magnetic fields may be able to significantly inhibit the decay of g-modes to turbulence.

\begin{figure}
\centerline{
\hspace{0.3cm}
\psfig{figure=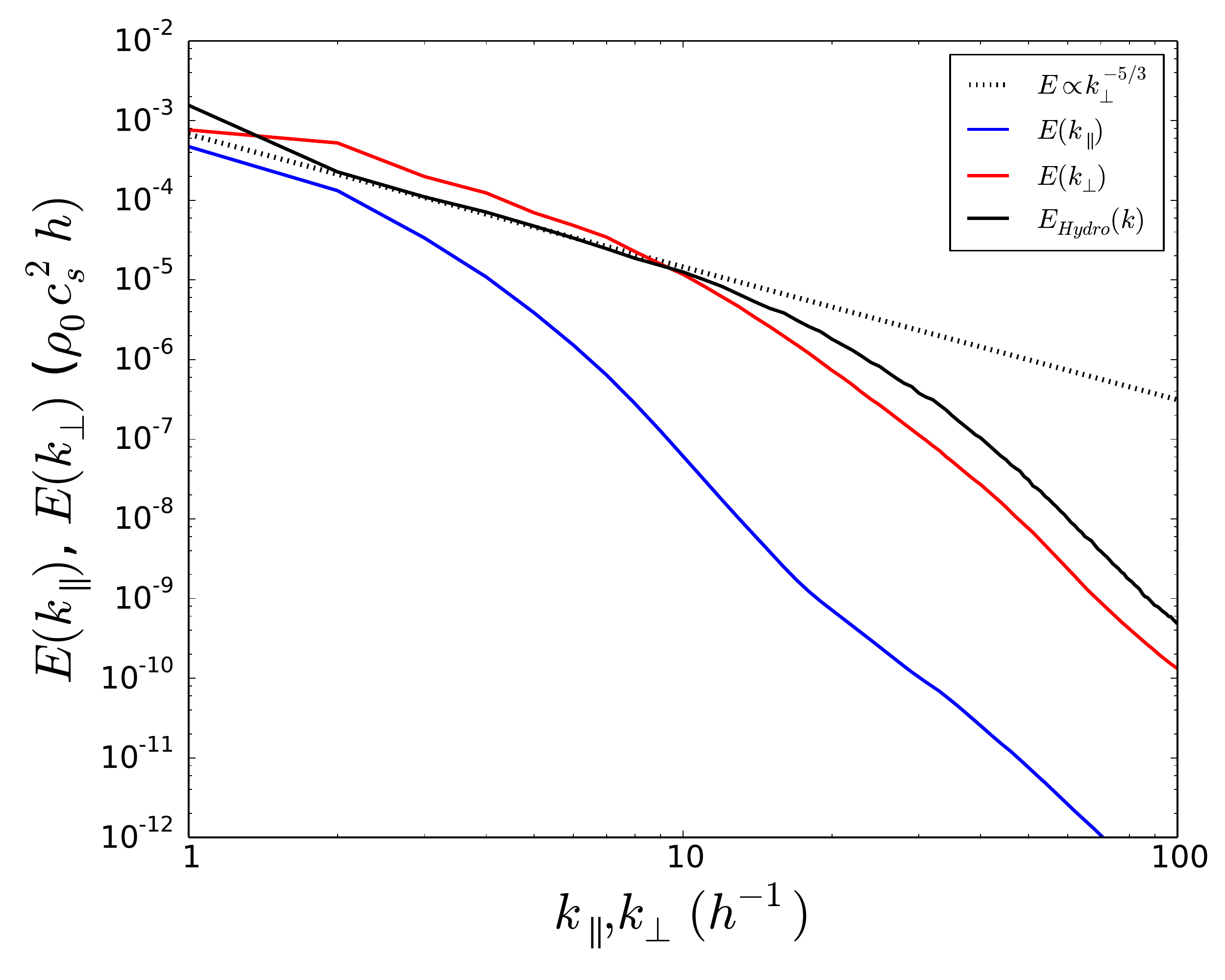,height=0.28\textheight}
}
\caption{The power spectrum of the unidirectional field simulation along the magnetic field lines (blue) shows a significantly softer spectrum than that across (red), indicating that the magnetic tension force acting along field lines acts to prevent the nonlinear breaking of g-modes and the subsequent transition to turbulence. Little energy reaches small scales, indicating a suppression of turbulence. Across the field, turbulence proceeds similar to the hydrodynamic control (black) with a Kolmogorov-like slope (dotted line). Spectra are weighted by the number of thermal bombs driving the hydrodynamic control run (27).}
\label{fig:anisotropic_power_spectrum}
\end{figure}

\subsection{Reflection of G-Modes}

Though we have addressed the majority of the observed dynamics in our simulations, there are a few notable features in the helical field simulation which are worth discussing. First of all, we see an inconsistency in the g-mode amplitude over the simulation time. This inconsistency is most prevalent in Figure 4b, in the ambient incompressible kinetic energy evolution. We note that the oscillating energy seems to form envelopes of increasing and decreasing amplitude. By eye, these oscillations appear to be consistent with a ``beating'' frequency between g-modes. 

The same interference effect is seen in the space-time plot of the Lagrangian tracer fluid which traces the evolution of g-modes (Figure 6b). Here, we see time periods where the g-mode seems to disappear at certain scale heights before recovering (e.g. $t$ = 300-400 $h/c_s$ for the tracer layer beginning at 4$h$). This feature in the g-mode amplitude appears to be consistent with the same beat frequency. 

How can two g-modes launched at different times interfere with each other in the helical field simulation, but not in the hydrodynamic control run? The answer appears to lie in a \textit{reflection} of the g-modes off of the magnetized atmosphere. Indeed, by analyzing the space-time evolution of the incompressible kinetic energy (Figure 10), we can analyze this effect. We see two arch-like features formed throughout the simulation, the first centered around 210 $h/c_s$, and the second centered around 350 $h/c_s$. The first strong g-mode is launched early on in the simulation, while the second appears to originate from just before the end of driving at 200 $h/c_s$. The first large g-mode reflects off of the magnetized atmosphere and propagates in the -$z$-direction. When the first, reflected g-mode intersects the second wave, the two waves interfere and produce a beating frequency. Because these waves are out of phase, the interference is destructive and appears to suppress the amplitude of g-modes within the interaction region.

The reflection of g-modes within a magnetized atmosphere is fully consistent with our original dispersion relation (Equation 1) and points to the well-known effect of g-mode trapping \citep{Balbus1990}. G-mode trapping occurs at a height where the g-mode frequency $\Omega_g(z) \approx N(z)$ where $N$ is the Brunt-V{\"a}is{\"a}l{\"a} frequency. In a purely hydrodynamic problem, this trapping occurs when $k_h$ = $k$ at some trapping height $z_0$. Despite choosing a trapping height far outside of our simulation domain ($z_0$ = 10$h$ in our simulations), MHD g-modes are able to be trapped. The reason is straightforward. Unlike the hydrodynamic g-mode, MHD g-modes possess a term analogous to that of a shear Alfv\`en wave, i.e. $v_A k_{\parallel}$. If the wavevector of a given g-mode has sufficient $k_{\parallel}$ such that $\Omega_g(z) \approx N(z)$, then a magnetized region with a density which would allow a purely hydrodynamic g-mode to propagate would become evanescent to MHD g-modes, and the incident wave would reflect. Within a helical field, there will always be a point where the wavevector lies along a field line, and thus the reflection of g-modes is inevitable within this configuration for our chosen strength of magnetic field.

The reflection/ trapping process presented here may be relevant in the ICM, preventing internal waves from propagating beyond the cool core. Even in situations where the entropy gradient would not trap a standard hydrodynamic g-mode, trapping will take place as long as the g-mode's wavevector has a sufficiently large component along the local magnetic field direction and the field is strong enough for the g-mode frequency to be comparable to the Brunt-V{\"a}is{\"a}l{\"a} frequency.

\begin{figure}
\centerline{
\hspace{0.3cm}
\psfig{figure=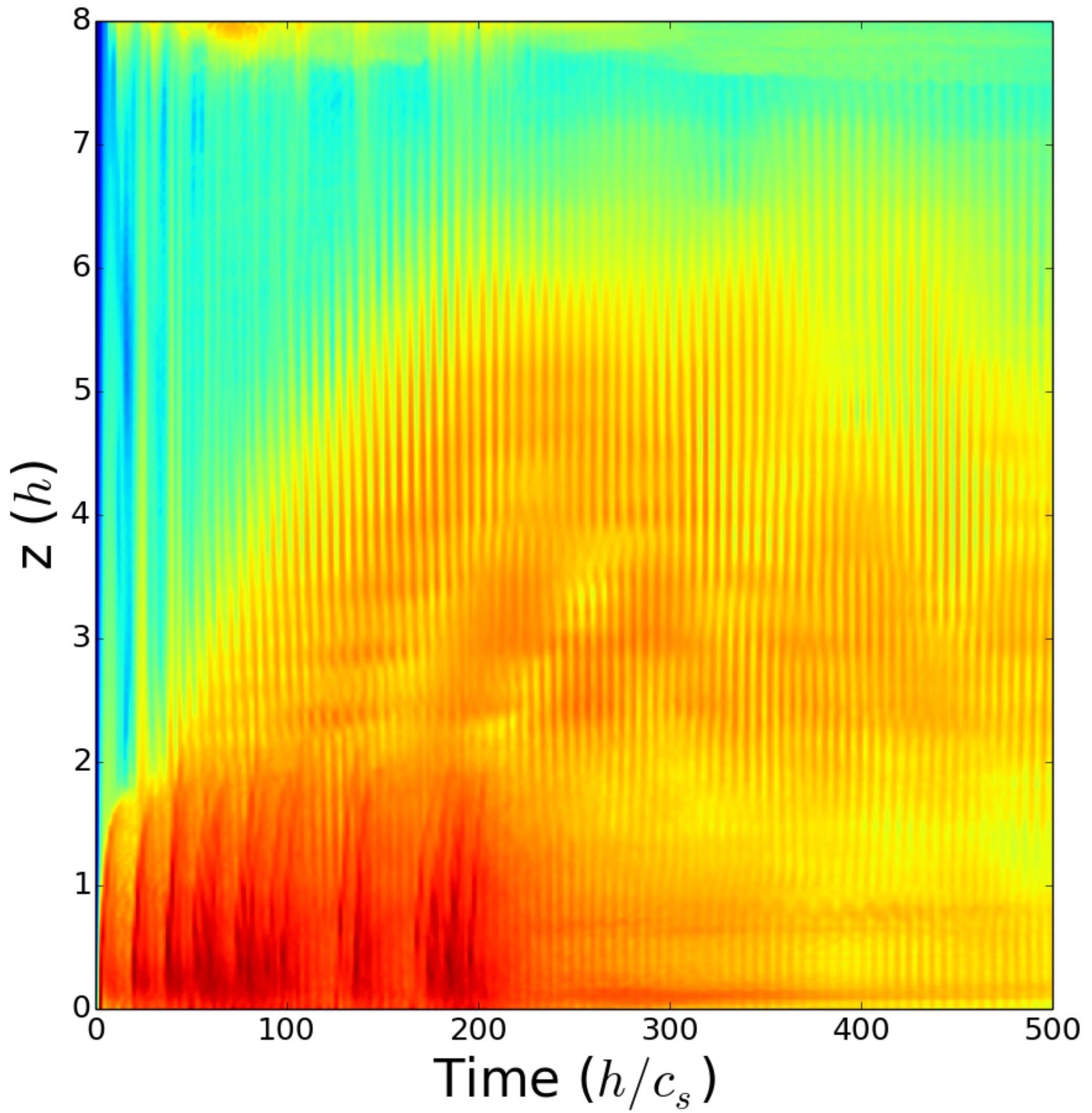,height=0.47\textwidth}
}
\caption{Space-time evolution of the incompressible kinetic energy of the helical field simulation. The incompressible kinetic energy displays two arch-like features. These features represent the rise of a g-mode oscillation and the subsequent ``fall'' of the energy following a reflection off of the magnetized atmosphere. In the time range where the two modes interfere, a beat-frequency is produced which suppresses the g-mode amplitude.}
\label{fig:g_mode_reflection}
\end{figure}

\section{Discussion}\label{discussion}

In this paper, we argue that a large-scale magnetic field cannot significantly increase the efficiency of g-mode driving through magnetic draping. In addition, we demonstrate how the magnetic tension force prevents the nonlinear decay of g-modes into turbulence, suppressing the transfer of energy to small scales, especially along the field lines. 

We argue that the inefficiency of driving turbulence within our simulations could be explained by 3 features: 1) the majority of kinetic energy is driven directly into compressible magnetosonic waves which rapidly leave our simulation domain, 2) incompressible kinetic energy is rapidly dissipated in the bubble debris region, and 3) the magnetic fields simply modify the magnetic RTI rather than fully suppressing this instability, leading to no significant increase in g-mode driving.

A secondary effect, magnetic tension, prevents the nonlinear transition of g-modes to turbulence. Thus, even if a helical magnetic field could preserve AGN-blown bubbles from the magnetic RTI and the KHI (as X-ray images clearly show well-preserved cavities far from the cluster center) and drive stronger g-modes, these g-modes would be preserved against the same instabilities by magnetic tension. Significantly stronger driving might be able to cause the nonlinear interaction time to become short enough as to induce breaking; however, this driving is still subject to the same loss mechanisms, namely magnetosonic waves and dissipation. Thus, the driving of turbulence remains inefficient.

While our argument appears clean, the argument contains a major flaw: it hinges on the validity of ideal MHD. Though we were able to motivate the use of ideal MHD in Section 2, the weak collisionality and high $\beta$ ICM is likely subject to pressure anisotropy-driven instabilities such as the firehose and mirror instabilities. The threshold criterion for a firehose-unstable plasma is given by \citep{Kunz2014},
\begin{equation}
	\frac{p_{\parallel} - p_{\bot}}{p} > \frac{2}{\beta}.
\end{equation}
For high $\beta$, the pressure anisotropy need not be large in order to reach the firehose threshold.

Pressure anisotropy-driven instabilities can have a number of effects. The firehose instability can act to negate magnetic tension, forcing the field lines to wave about like an uncontrolled firehose. This negation of magnetic tension may allow our g-modes (or gravito-Alfv\`en waves depending on how one views the dispersion relation) to disrupt and transition into nonlinear turbulence, heating the cluster \citep{Squire2016}. Furthermore, the firehose instability may suppress the tension which cuts our turbulent cascade at small scales, allowing turbulence to proceed in an almost hydrodynamic manner. 

If we include the firehose instability however, we must take into account the mirror instability, driven by a dominance of perpendicular pressure to parallel pressure. This instability is largely activated by the amplification of magnetic field, perhaps through stretching in the magnetic draping process. In this way, one could imagine a series of magnetic mirrors forming all along our draped magnetic field, eventually reconnecting and breaking up the field, fully suppressing the magnetic draping process. Future work could approach this problem through the inclusion of an anisotropic (Braginskii) viscosity; however, without significant amplification of g-modes, there is still insufficient energy to offset radiative cooling present within incompressible modes in the ICM.

Thermal conduction effects may also play a role. Our simulations include no thermal conduction in order to focus explicitly on turbulent driving. If a temperature gradient is included in the presence of a magnetized atmosphere, the atmosphere becomes subject to the magnetothermal instability (MTI) (see \cite{Balbus2000}) and heat-flux buoyancy instability (HBI) (see \cite{Quataert2008}), each of which can drive turbulence. However, we note that kinetic instabilities such as mirror and Whistler wave instabilities may serve to strongly suppress thermal conduction \citep{RobergClark2016, RobergClark2018}. 

Finally, even if bubbles can be fully preserved \citep{Zhang2018}, turbulent heating of the Perseus Cluster has a major problem: the turbulence cannot propagate to the cooling radius of the cluster fast enough. This result, discussed in \cite{Fabian2017} and extended to other clusters in Bambic et al. (2018b, in prep.) may indicate that another mechanism is responsible for heating galaxy clusters. Fabian et al. suggest sound waves. Indeed, within our simulations, compressible modes carry away the majority of the kinetic energy (10\% of the injected energy in Simulations 1 and 2). Thus, the bubble inflation process seems to be a viable means of transferring sufficient energy to magnetosonic waves in order to offset radiative cooling. Most theories of sound wave heating currently rest on a hydrodynamic description of dissipation, invoking a variable viscosity as a means of dissipating the wave \citep{Ruszkowski2004}. Because the ICM viscosity is not well defined \citep{Su2017}, more effort within the plasma astrophysics community is required to elucidate the detailed physics of compressible wave dissipation within a weakly collisional plasma.
In addition, cosmic ray streaming may provide a viable means of heating clusters. As is indicated by work from \cite{Ruszkowski2017}, cosmic rays provide a self-regulating means of feedback. Because current simulations require a fluid treatment of cosmic rays, more efforts will be required within plasma kinetics in order to understand this mechanism.

Due to the complexity and diversity of plasma phenomena likely to operate within the ICM, we encourage future theoretical work to address these issues separately so as to properly isolate the physics relevant to each process. 

\section{Conclusion}\label{conclusion}
In this paper, we argue that magnetic draping is unable to noticeably amplify the production of g-modes or internal waves within the ICM plasma. When a magnetic field acts in only one direction along an AGN-driven bubble, the bubble-ICM interface is subject to a modified form of the Rayleigh-Taylor instability, what we refer to in this work as the magnetic RTI. The magnetic RTI appears to be partially suppressed by a helical magnetic field; however, the instability continues in a modified fashion, producing twisted ribbon-like structures in the bubbles rather than the traditional vortex rings or sheets (as is the case for field in only one direction). The challenge of amplifying g-modes is heightened by loss mechanisms operating in the magnetized atmosphere: energy from the initial driving is siphoned off into stretching the magnetic field, buoyantly lifting the atmosphere, and heating the bubble debris region. Finally, the driving of turbulence itself is highly inefficient (0.014\% efficiency), with magnetic tension preventing the nonlinear decay of g-modes to turbulence and suppressing that turbulence at the smallest scales. While the effect of magnetic tension can be negated through a more complicated treatment of the high $\beta$ ICM plasma, this treatment may prove to negate the magnetic draping process entirely, resulting in the same inefficiency with AGN-driven turbulence identified by RBS.

In the absence of a clear explanation for AGN-driven turbulence, other physics could be responsible for the observed feedback. More effort is required to elucidate the weakly-collisional, high $\beta$ plasma physics relevant to the ICM, including mechanisms for sound wave dissipation and cosmic ray streaming. Overall, we encourage the plasma astrophysics community to continue to isolate the physics relevant to the ICM plasma as we have done in this paper, and systematically analyze mechanisms for thermalizing the AGN energy. We hope that our work points the community toward a better understanding of ICM plasma physics and moves us one step closer toward understanding the deep connection between microscale plasma physics and macroscale AGN feedback.

\acknowledgements

CJB would like to thank William Dorland of the University of Maryland as well as Steven Balbus and Alexander Schekochihin of the University of Oxford for a number of helpful discussions leading up to the publication of this work. In addition, CJB is grateful for the financial support of NSF grant AST1333514. CSR and BJM thank the support of the National Science Foundation under the Theory and Computational Astrophysics Network grant AST1333514.  CSR also thanks the support of NASA under the Astrophysics Theory Program grant  NNX17AG27G. We would also like to thank an anonymous referee for comments and suggestions prior to the publication of this work. The simulations presented in this paper were performed on the Deepthought2 cluster, maintained and supported by the Division of Information Technology at the University of Maryland, College Park.


\begin{thebibliography}{}

\bibitem[\protect\astroncite{{Balbus}}{2000}]{Balbus2000}
{Balbus}, S.~A.,  2000, \apj, 534, 420

\bibitem[\protect\astroncite{{Balbus} \& {Reynolds}}{2008}]{Balbus2008}
{Balbus}, S.~A., \& {Reynolds}, C.~S.,  2008, \apjl, 681, L65

\bibitem[\protect\astroncite{{Balbus} \& {Reynolds}}{2010}]{Balbus2010}
{Balbus}, S.~A., \& {Reynolds}, C.~S.,  2010, \apjl, 720, L97

\bibitem[\protect\astroncite{{Balbus} \& {Soker}}{1990}]{Balbus1990}
{Balbus}, S.~A., \& {Soker}, N.,  1990, \apj, 357, 353

\bibitem[\protect\astroncite{{Binney} \& {Cowie}}{1981}]{Binney1981}
{Binney}, J., \& {Cowie}, L.~L.,  1981, \apj, 247, 464

\bibitem[\protect\astroncite{{Bogdanovi{\'c}} et~al.}{2009}]{Bogdanovic2009}
{Bogdanovi{\'c}}, T., {Reynolds}, C.~S., {Balbus}, S.~A., \& {Parrish}, I.~J.,
  2009, \apj, 704, 211

\bibitem[\protect\astroncite{{Bonafede} et~al.}{2010}]{Bonafede2010}
{Bonafede}, A., {Feretti}, L., {Murgia}, M., {Govoni}, F., {Giovannini}, G.,
  {Dallacasa}, D., {Dolag}, K., \& {Taylor}, G.~B.,  2010, \aap, 513, A30

\bibitem[\protect\astroncite{{Bourne} \& {Sijacki}}{2017}]{Bourne2017}
{Bourne}, M.~A., \& {Sijacki}, D.,  2017, ArXiv e-prints

\bibitem[\protect\astroncite{Brouillette}{2002}]{Brouillette2002}
Brouillette, M.,  2002, Annual Review of Fluid Mechanics, 34, 445

\bibitem[\protect\astroncite{{Burns}}{1990}]{Burns1990}
{Burns}, J.~O.,  1990, \aj, 99, 14

\bibitem[\protect\astroncite{{Chandrasekhar}}{1961}]{Chandrasekhar1961}
{Chandrasekhar}, S.,  1961,
\newblock {Hydrodynamic and hydromagnetic stability}

\bibitem[\protect\astroncite{{Churazov} et~al.}{2000}]{Churazov2000}
{Churazov}, E., {Forman}, W., {Jones}, C., \& {B{\"o}hringer}, H.,  2000, \aap,
  356, 788

\bibitem[\protect\astroncite{{Churazov} et~al.}{2002}]{Churazov2002}
{Churazov}, E., {Sunyaev}, R., {Forman}, W., \& {B{\"o}hringer}, H.,  2002,
  \mnras, 332, 729

\bibitem[\protect\astroncite{{Dursi} \& {Pfrommer}}{2008}]{Dursi2008}
{Dursi}, L.~J., \& {Pfrommer}, C.,  2008, \apj, 677, 993

\bibitem[\protect\astroncite{{Evans} \& {Hawley}}{1988}]{Evans1988}
{Evans}, C.~R., \& {Hawley}, J.~F.,  1988, \apj, 332, 659

\bibitem[\protect\astroncite{{Fabian}}{1994}]{Fabian1994}
{Fabian}, A.~C.,  1994, \araa, 32, 277

\bibitem[\protect\astroncite{{Fabian}}{2012}]{Fabian2012}
{Fabian}, A.~C.,  2012, \araa, 50, 455

\bibitem[\protect\astroncite{{Fabian} et~al.}{2000}]{Fabian2000}
{Fabian}, A.~C., et~al., 2000, \mnras, 318, L65

\bibitem[\protect\astroncite{{Fabian} et~al.}{2006}]{Fabian2006}
{Fabian}, A.~C., {Sanders}, J.~S., {Taylor}, G.~B., {Allen}, S.~W., {Crawford},
  C.~S., {Johnstone}, R.~M., \& {Iwasawa}, K.,  2006, \mnras, 366, 417

\bibitem[\protect\astroncite{{Fabian} et~al.}{2017}]{Fabian2017}
{Fabian}, A.~C., {Walker}, S.~A., {Russell}, H.~R., {Pinto}, C., {Sanders},
  J.~S., \& {Reynolds}, C.~S.,  2017, \mnras, 464, L1

\bibitem[\protect\astroncite{{Felten} et~al.}{1966}]{Felten1966}
{Felten}, J.~E., {Gould}, R.~J., {Stein}, W.~A., \& {Woolf}, N.~J.,  1966,
  \apj, 146, 955

\bibitem[\protect\astroncite{{Forman} et~al.}{2017}]{Forman2017}
{Forman}, W., {Churazov}, E., {Jones}, C., {Heinz}, S., {Kraft}, R., \&
  {Vikhlinin}, A.,  2017, \apj, 844, 122

\bibitem[\protect\astroncite{{Friedman}, {Heinz} \&
  {Churazov}}{2012}]{Friedman2012}
{Friedman}, S.~H., {Heinz}, S., \& {Churazov}, E.,  2012, \apj, 746, 112

\bibitem[\protect\astroncite{{Ghosh} et~al.}{2009}]{Ghosh2009}
{Ghosh}, S., {Thomson}, D.~J., {Matthaeus}, W.~H., \& {Lanzerotti}, L.~J.,
  2009, Journal of Geophysical Research (Space Physics), 114, A08106

\bibitem[\protect\astroncite{{Goldreich} \& {Sridhar}}{1995}]{Goldreich1995}
{Goldreich}, P., \& {Sridhar}, S.,  1995, \apj, 438, 763

\bibitem[\protect\astroncite{{Hillel} \& {Soker}}{2016}]{Hillel2016}
{Hillel}, S., \& {Soker}, N.,  2016, \mnras, 455, 2139

\bibitem[\protect\astroncite{{Hillel} \& {Soker}}{2017}]{Hillel2017}
{Hillel}, S., \& {Soker}, N.,  2017, ArXiv e-prints

\bibitem[\protect\astroncite{{Hitomi Collaboration} et~al.}{2016}]{Hitomi2016}
{Hitomi Collaboration}et~al., 2016, \nat, 535, 117

\bibitem[\protect\astroncite{{Howes} \& {Nielson}}{2013}]{Howes2013}
{Howes}, G.~G., \& {Nielson}, K.~D.,  2013, Physics of Plasmas, 20, 072302

\bibitem[\protect\astroncite{{Jeans}}{1902}]{Jeans}
{Jeans}, J.~H.,  1902, Philosophical Transactions of the Royal Society of
  London Series A, 199, 1

\bibitem[\protect\astroncite{{Jones} et~al.}{2002}]{Jones2002}
{Jones}, C., {Forman}, W., {Vikhlinin}, A., {Markevitch}, M., {David}, L.,
  {Warmflash}, A., {Murray}, S., \& {Nulsen}, P.~E.~J.,  2002, \apjl, 567, L115

\bibitem[\protect\astroncite{{Kim} \& {Narayan}}{2003}]{Kim2003}
{Kim}, W.-T., \& {Narayan}, R.,  2003, \apj, 596, 889

\bibitem[\protect\astroncite{{Kunz}, {Schekochihin} \&
  {Stone}}{2014}]{Kunz2014}
{Kunz}, M.~W., {Schekochihin}, A.~A., \& {Stone}, J.~M.,  2014, Physical Review
  Letters, 112, 205003

\bibitem[\protect\astroncite{Meshkov}{1969}]{Meshkov1969}
Meshkov, E.~E.,  1969, Fluid Dynamics, 4, 101

\bibitem[\protect\astroncite{{Mignone} et~al.}{2007}]{Mignone2007}
{Mignone}, A., {Bodo}, G., {Massaglia}, S., {Matsakos}, T., {Tesileanu}, O.,
  {Zanni}, C., \& {Ferrari}, A.,  2007, \apjs, 170, 228

\bibitem[\protect\astroncite{{Million} et~al.}{2010}]{Million2010}
{Million}, E.~T., {Werner}, N., {Simionescu}, A., {Allen}, S.~W., {Nulsen},
  P.~E.~J., {Fabian}, A.~C., {B{\"o}hringer}, H., \& {Sanders}, J.~S.,  2010,
  \mnras, 407, 2046

\bibitem[\protect\astroncite{{Narayan} \& {Medvedev}}{2001}]{Narayan2001}
{Narayan}, R., \& {Medvedev}, M.~V.,  2001, \apjl, 562, L129

\bibitem[\protect\astroncite{{Parrish} et~al.}{2012}]{Parrish2012}
{Parrish}, I.~J., {McCourt}, M., {Quataert}, E., \& {Sharma}, P.,  2012,
  \mnras, 419, L29

\bibitem[\protect\astroncite{{Peterson} \& {Fabian}}{2006}]{Peterson2006}
{Peterson}, J.~R., \& {Fabian}, A.~C.,  2006, \physrep, 427, 1

\bibitem[\protect\astroncite{{Quataert}}{2008}]{Quataert2008}
{Quataert}, E.,  2008, \apj, 673, 758

\bibitem[\protect\astroncite{{Randall} et~al.}{2015}]{Randall2015}
{Randall}, S.~W., et~al., 2015, \apj, 805, 112

\bibitem[\protect\astroncite{{Reynolds}, {Balbus} \&
  {Schekochihin}}{2015}]{RBS15}
{Reynolds}, C.~S., {Balbus}, S.~A., \& {Schekochihin}, A.~A.,  2015, \apj, 815,
  41

\bibitem[\protect\astroncite{{Reynolds}, {Heinz} \&
  {Begelman}}{2002}]{Reynolds2002}
{Reynolds}, C.~S., {Heinz}, S., \& {Begelman}, M.~C.,  2002, \mnras, 332, 271

\bibitem[\protect\astroncite{Richtmyer}{1960}]{Richtmyer1960}
Richtmyer, R.~D.,  1960, Communications on Pure and Applied Mathematics, 13,
  297

\bibitem[\protect\astroncite{{Roberg-Clark} et~al.}{2016}]{RobergClark2016}
{Roberg-Clark}, G.~T., {Drake}, J.~F., {Reynolds}, C.~S., \& {Swisdak}, M.,
  2016, \apjl, 830, L9

\bibitem[\protect\astroncite{{Roberg-Clark} et~al.}{2017}]{RobergClark2018}
{Roberg-Clark}, G.~T., {Drake}, J.~F., {Reynolds}, C.~S., \& {Swisdak}, M.,
  2017, ArXiv e-prints

\bibitem[\protect\astroncite{{Ruszkowski} \& {Begelman}}{2002}]{Ruszkowski2002}
{Ruszkowski}, M., \& {Begelman}, M.~C.,  2002, \apj, 581, 223

\bibitem[\protect\astroncite{{Ruszkowski}, {Br{\"u}ggen} \&
  {Begelman}}{2004}]{Ruszkowski2004}
{Ruszkowski}, M., {Br{\"u}ggen}, M., \& {Begelman}, M.~C.,  2004, \apj, 611,
  158

\bibitem[\protect\astroncite{{Ruszkowski} et~al.}{2007}]{Ruszkowski2007}
{Ruszkowski}, M., {En{\ss}lin}, T.~A., {Br{\"u}ggen}, M., {Heinz}, S., \&
  {Pfrommer}, C.,  2007, \mnras, 378, 662

\bibitem[\protect\astroncite{{Ruszkowski}, {Yang} \&
  {Reynolds}}{2017}]{Ruszkowski2017}
{Ruszkowski}, M., {Yang}, H.-Y.~K., \& {Reynolds}, C.~S.,  2017, \apj, 844, 13

\bibitem[\protect\astroncite{{Schekochihin} et~al.}{2009}]{Schekochihin2009}
{Schekochihin}, A.~A., {Cowley}, S.~C., {Dorland}, W., {Hammett}, G.~W.,
  {Howes}, G.~G., {Quataert}, E., \& {Tatsuno}, T.,  2009, \apjs, 182, 310

\bibitem[\protect\astroncite{{Schwarzschild}}{1958}]{Schwarzschild}
{Schwarzschild}, M.,  1958,
\newblock {Structure and evolution of the stars.}

\bibitem[\protect\astroncite{{Squire}, {Quataert} \&
  {Schekochihin}}{2016}]{Squire2016}
{Squire}, J., {Quataert}, E., \& {Schekochihin}, A.~A.,  2016, \apjl, 830, L25

\bibitem[\protect\astroncite{{Stone} \& {Gardiner}}{2007}]{Stone2007}
{Stone}, J.~M., \& {Gardiner}, T.,  2007, \apj, 671, 1726

\bibitem[\protect\astroncite{{Su} et~al.}{2017}]{Su2017}
{Su}, Y., et~al., 2017, \apj, 834, 74

\bibitem[\protect\astroncite{{Tang} \& {Churazov}}{2017}]{Tang2017}
{Tang}, X., \& {Churazov}, E.,  2017, \mnras, 468, 3516

\bibitem[\protect\astroncite{{Toro}, {Spruce} \& {Speares}}{1994}]{Toro1994}
{Toro}, E.~F., {Spruce}, M., \& {Speares}, W.,  1994, Shock Waves, 4, 25

\bibitem[\protect\astroncite{{Voigt} \& {Fabian}}{2004}]{Voigt2004}
{Voigt}, L.~M., \& {Fabian}, A.~C.,  2004, \mnras, 347, 1130

\bibitem[\protect\astroncite{{Weinberger} et~al.}{2017}]{Weinberger2017}
{Weinberger}, R., {Ehlert}, K., {Pfrommer}, C., {Pakmor}, R., \& {Springel},
  V.,  2017, ArXiv e-prints

\bibitem[\protect\astroncite{{Yang} \& {Reynolds}}{2016a}]{Yang2016}
{Yang}, H.-Y.~K., \& {Reynolds}, C.~S.,  2016a, \apj, 829, 90

\bibitem[\protect\astroncite{{Yang} \& {Reynolds}}{2016b}]{Yang2016a}
{Yang}, H.-Y.~K., \& {Reynolds}, C.~S.,  2016b, \apj, 818, 181

\bibitem[\protect\astroncite{{Zakamska} \& {Narayan}}{2003}]{Zakamska2003}
{Zakamska}, N.~L., \& {Narayan}, R.,  2003, \apj, 582, 162

\bibitem[\protect\astroncite{{Zhang}, {Churazov} \&
  {Schekochihin}}{2018}]{Zhang2018}
{Zhang}, C., {Churazov}, E., \& {Schekochihin}, A.~A.,  2018, ArXiv e-prints

\bibitem[\protect\astroncite{{Zhuravleva} et~al.}{2017}]{Zhuravleva2017}
{Zhuravleva}, I., {Allen}, S.~W., {Mantz}, A.~B., \& {Werner}, N.,  2017, ArXiv
  e-prints

\bibitem[\protect\astroncite{{Zhuravleva} et~al.}{2016}]{Zhuravleva2016}
{Zhuravleva}, I., et~al., 2016, \mnras, 458, 2902

\bibitem[\protect\astroncite{{Zhuravleva} et~al.}{2014}]{Zhuravleva2014}
{Zhuravleva}, I., et~al., 2014, \nat, 515, 85

\bibitem[\protect\astroncite{{ZuHone}, {Markevitch} \&
  {Lee}}{2011}]{ZuHone2011}
{ZuHone}, J.~A., {Markevitch}, M., \& {Lee}, D.,  2011, \apj, 743, 16

\end{thebibliography}

\end{document}